%% file: main.tex
\documentclass{sigchi}

\CopyrightYear{2020}
\setcopyright{acmlicensed}
\doi{https://doi.org/10.1145/3313831.3376735}
\isbn{978-1-4503-6708-0/20/04}
\conferenceinfo{CHI '20;}{April 25--30, 2020; Honolulu, HI, USA}
\acmPrice{\$15.00}

\toappear{\scriptsize Permission to make digital or hard copies of all or part of this work for personal or classroom use is granted without fee provided that copies are not made or distributed for profit or commercial advantage and that copies bear this notice and the full citation on the first page. Copyrights for components of this work owned by others than ACM must be honored. Abstracting with credit is permitted. To copy otherwise, or republish, to post on servers or to redistribute to lists, requires prior specific permission and/or a fee. Request permissions from \href{mailto:Permissions@acm.org}{Permissions@acm.org} \\
{\emph{CHI '20}, April 25--30, 2020, Honolulu, HI, USA. } \\
\copyright~2020 Association for Computing Machinery. \\
ACM ISBN 978-1-4503-6708-0/20/04\ ...\$15.00. \\
\url{https://doi.org/10.1145/3313831.3376735}}

\usepackage{balance}       
\usepackage{graphics}      
\usepackage[T1]{fontenc}   
\usepackage{txfonts}
\usepackage{mathptmx}
\usepackage[pdflang={en-US},pdftex]{hyperref}
\usepackage{color}
\usepackage{booktabs}
\usepackage{textcomp}
\usepackage{gensymb}
\usepackage[english]{babel}
\usepackage{afterpage}
\usepackage{cite}

\usepackage{amsthm}

\usepackage{microtype}        
\usepackage{ccicons}          

\def\plaintitle{Optimal Sensor Position for a Computer Mouse}

\def\emptyauthor{}
\def\plainkeywords{Computer; mouse; sensor position; pointing performance; virtual sensor position; optimization}

\makeatletter
\def\url@leostyle{%
  \@ifundefined{selectfont}{
    \def\UrlFont{\sf}
  }{
    \def\UrlFont{\small\bf\ttfamily}
  }}
\makeatother
\def\pprw{8.5in}
\def\pprh{11in}

\definecolor{linkColor}{RGB}{6,125,233}
\hypersetup{%
  pdftitle={\plaintitle},
  pdfauthor={\emptyauthor},
  pdfkeywords={\plainkeywords},
  pdfdisplaydoctitle=true, 
  bookmarksnumbered,
  pdfstartview={FitH},
  colorlinks,
  citecolor=black,
  filecolor=black,
  linkcolor=black,
  urlcolor=linkColor,
  breaklinks=true,
  hypertexnames=false
}

\begin{document}

\title{\plaintitle}

\numberofauthors{1}
\author{
	\alignauthor{Sunjun Kim\textsuperscript{1,3} \hspace{1cm} Byungjoo Lee\textsuperscript{2} \hspace{1cm} Thomas van Gemert\textsuperscript{1} \hspace{1cm} Antti Oulasvirta\textsuperscript{1}\\
		\affaddr{\textsuperscript{1}Aalto University, Espoo, Finland \hspace{0.6cm}
			\textsuperscript{2}KAIST, Daejeon, Republic of Korea}\\
			\textsuperscript{3}DGIST, Daegu, Republic of Korea\\
		\email{sunjun.kim@aalto.fi},
		\email{byungjoo.lee@kaist.ac.kr},
		\email{thomas.vangemert@aalto.fi},
		\email{antti.oulasvirta@aalto.fi}
	}
}
\maketitle

\begin{abstract}
\input{abstract}
\end{abstract}




\section{Introduction}
\input{sec_introduction}


 

\section{Related Work}  
\input{sec_relwork}

\section{A Variable-Sensor-Position Mouse Device}
\input{sec_virtual_sensor}

\section{Study: Effect on Pointing Performance}
\input{sec_experiment}

\section{Calibration and Software-Side Tuning}
\input{sec_calibration}


\section{Summary and Discussion}
\input{sec_discussion.tex}

\balance{}

\bibliographystyle{SIGCHI-Reference-Format}
\bibliography{ref}

\end{document}

%% file: abstract.tex
Computer mice have their displacement sensors in various locations (center, front, and rear). However, there has been little research into the effects of sensor position or on engineering approaches to exploit it. 
This paper first discusses the mechanisms via which sensor position affects mouse movement and reports the results from a study of a pointing task in which the sensor position was systematically varied. 
Placing the sensor in the center turned out to be the best compromise: 
improvements over front and rear were in the 11--14\% range for throughput and 20--23\% for path deviation. 
However, users varied in their personal optima. 
Accordingly, variable-sensor-position mice are then presented, with a demonstration that high accuracy can be achieved with two static optical sensors. 
A virtual sensor model is described that allows software-side repositioning of the sensor. 
Individual-specific calibration should yield an added 4\% improvement in throughput over the default center position.

%% file: sec_introduction.tex
In the wake of Douglas Engelbart's Mother of All Demos,
the computer mouse has grown into one of the most engineered input devices \cite{engelbart1970xy}.
Besides proprietary work, there has been published research on almost every factor imaginable:
the shape of the device \cite{Chen2007ey,dehghan2013designing,gustafsson2003computer,Hedge2010ep,houwink2009providing,keir1999effects,lee2007alternative,odell2007evaluation,quemelo2013biomechanics},
its weight \cite{chen2012weight},
the control-to-display (CD) gain function \cite{casiez2011no, casiez08, anglemouse}, the resolution of the displacement sensor \cite{aceituno2013low, roussel2012subpixel, casiez08, shoemaker2012two}, and many more. 
Yet it is fascinating to observe that for \emph{one} factor, the placement of the sensor in the bottom of the device,
the proper examination has been overlooked in academic research.
This factor turns out to have a strong but exploitable effect on users' pointing performance.

\begin{figure}[t!]
\centering
  \includegraphics[width=0.8\columnwidth]{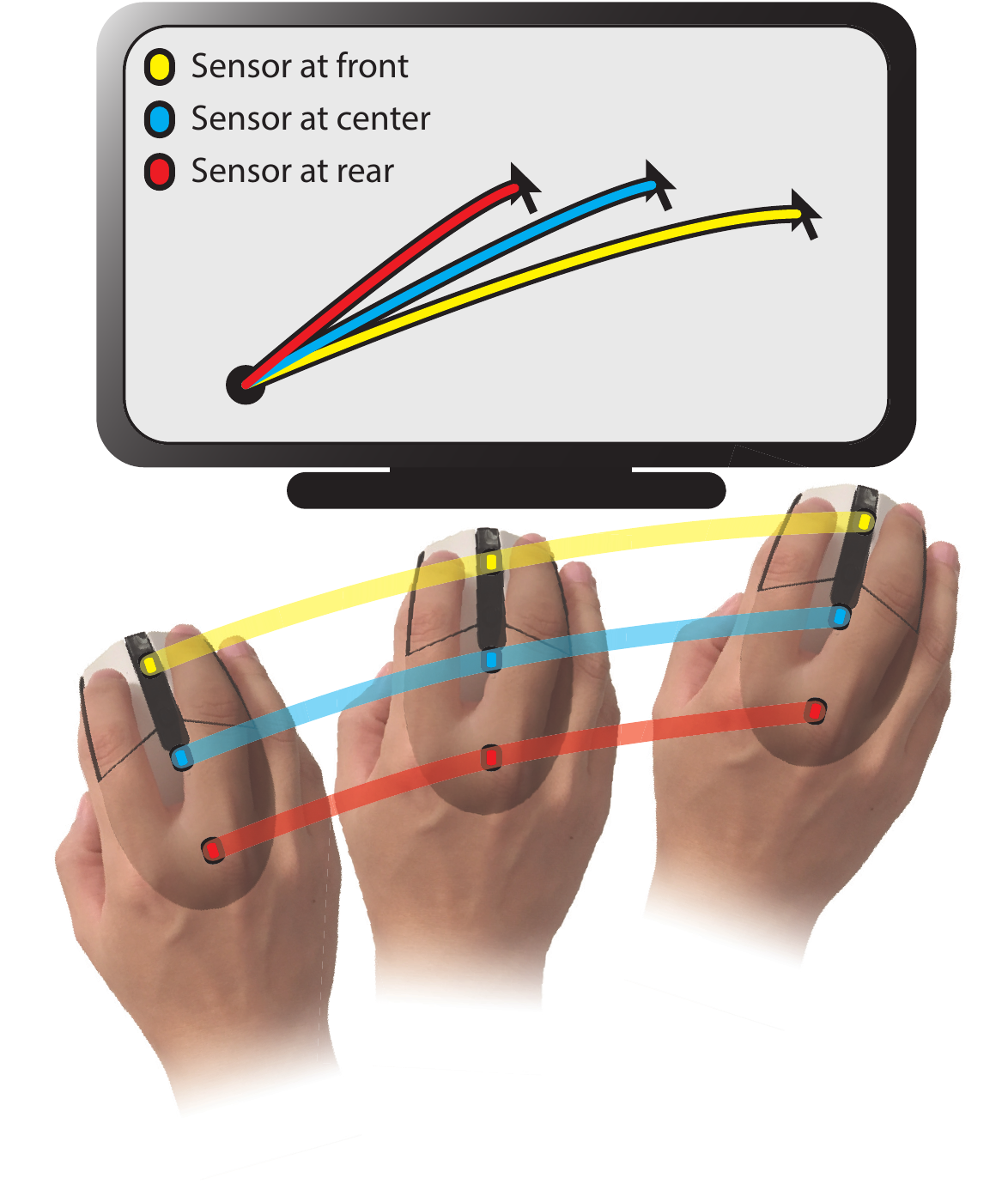}\vspace{-2.1em}
  \caption{
When a user operates a mouse, the mouse moves and rotates, and the resulting trajectory is affected by the position of the displacement sensor. 
 When it is closer to the front, the increased radius of the arc results in the larger horizontal cursor displacement.
Vertical cursor displacement stays the same irrespective of sensor position. 
 }~\label{fig:problem}
\vspace{-3em}
\end{figure}

To understand why sensor position matters, 
we need to revisit a misconception about mouse movement.
A mouse is a transducer that converts a planar movement into a two-dimensional (2D) cursor displacement. 
We might thus readily believe that there is a one-to-one mapping between the physical and cursor movement, 
or at least that the mapping is modulated by the CD gain function. 
That is not, in fact, the full picture. 
The position of the displacement sensor has a covert effect on cursor movement. 
The planar motion of a mouse, when held in hand, 
is produced as a combination of two-dimensional translation and one-dimensional rotation. 
Though users may \emph{perceive} translational movement of the cursor, 
joint rotation adds inadvertent rotational motion of the mouse body. 
The wrist joint rotates about 15--25 degrees (radial--ulnar deviation) during mouse pointing \cite{burgess1999wrist, chen2012weight, jensen1998job, wahlstrom2000differences}. 
Also, the shoulder joint rotates, producing rotation around the elbow joint (mediolateral deviation).
Firstly, the sensor ignores rotation in motion, which distorts the cursor trajectory \cite{Lee2015}. 
More interestingly, when the displacement sensor is placed further from or closer to the pivot of rotation, 
different paths result, as shown in Figure \ref{fig:problem}.
The differences can be large.
Our study revealed that a sensor positioned at the front produced almost twice the horizontal cursor displacement than one at the rear did (this is discussed further on).

The discussion surrounding sensor position is riddled with unproven hypotheses and misconceptions.
There is only a single academic paper on the topic, from 1989: 
Verplank and Oliver compared the task completion time in a maze-tracing task ($N=5$) with three mice prototypes, varying their shape and the position of ball sensors \cite{verplank1989microsoft}. 
They concluded that a sensor position toward the front is favorable.
The designers of the Microsoft Mouse took this finding to indicate ``a dramatic (performance) advantage in moving the ball to the front of the mouse'' \cite{moggridge2007designing},
explaining the advantage as being due to ``increasing the apparent moment of inertia as the mouse is pivoted from [the] elbow or heel of the hand'' \cite{verplank1989microsoft}. 
However, this conclusion is potentially flawed as well as outdated.
Their study confounded the position of the sensor with the shape of the mouse, which we now know affects performance. Moreover, no statistical analysis was reported.
Secondly, they used a low-CPI ball mouse of that era. 
A modern optical mouse sensor operates like a camera with high speed but low resolution. 
It computes translational displacement by calculating the cross-correlation between successive images.
Importantly, resolution, measured in Counts per Inch, or CPI,
has improved dramatically. 
While the mice of the 1980s have 100--400 CPI, an off-the-shelf mouse in 2019 exceeds 800 CPI, and a high-grade sensor can easily reach above 10K CPI, which is far beyond the human limit \cite{berard2011humanLimit}.  

\begin{figure}[t!]
	\centering
	\includegraphics[width=0.999\columnwidth]{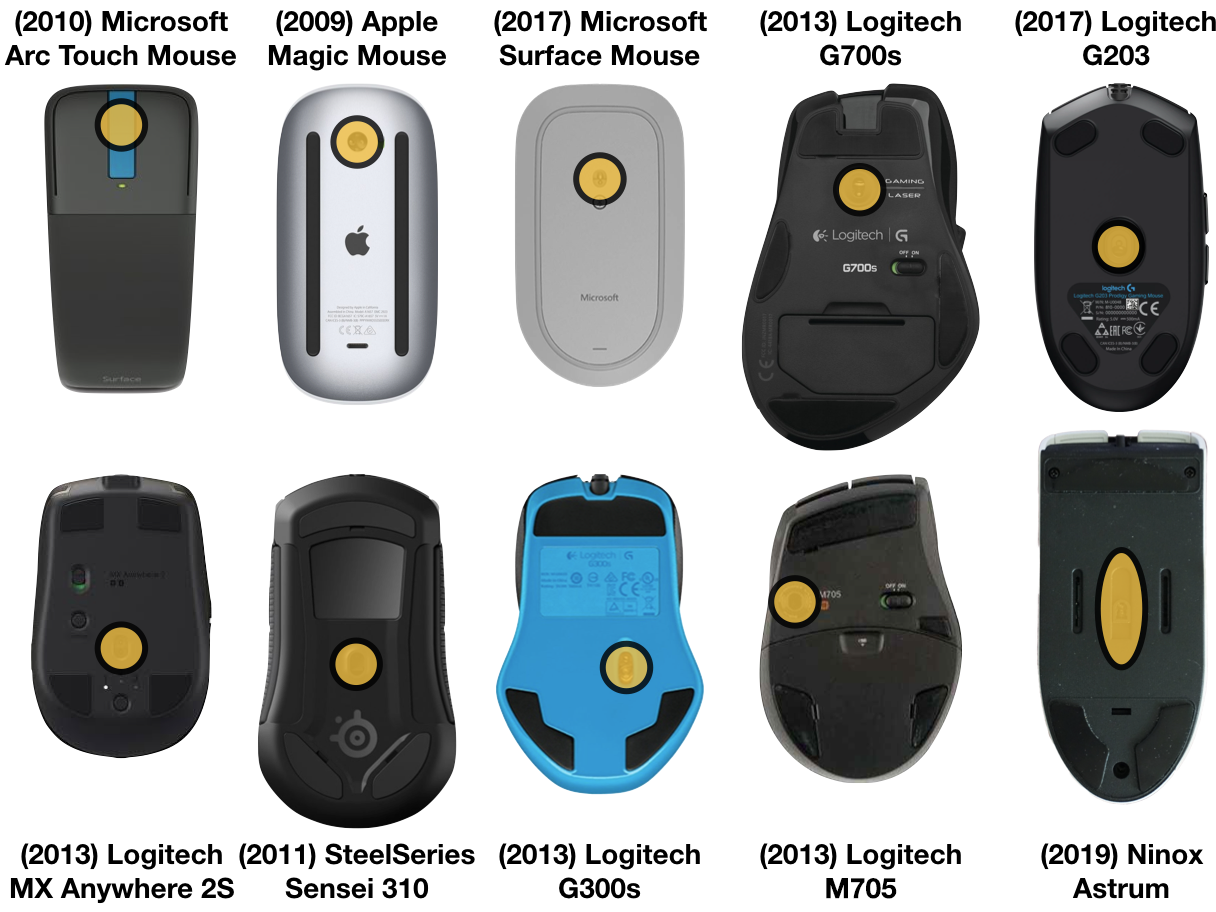}
	\caption{Commercial mice (shown to scale) exhibit a wide distribution of displacement sensor designs. Note that \textit{Ninox Astrum} is equipped with a slider for a variable sensor position.}~\label{fig:sensor_positions}
	\vspace{-1em}
\end{figure}

Within internet communities,
two different hypotheses exist \cite{myth_esr,myth_oc1,myth_geekhack,myth_reddit}. 
According to the first, a sensor beneath the index finger performs best, because it reacts more swiftly (due to larger radius), and
experience can be transferred from precise pen and touch manipulations. 
The second hypothesis holds that the center is best,
because an object can be more easily manipulated, supposedly, when force is exerted at its center of gravity. 
In any case, both hypotheses have not yet been verified.
Manufacturers have taken very different positions on the topic, as Figure \ref{fig:sensor_positions} attests. 
The center position seems to dominate, but the front and back also are represented.

To the best of our knowledge, this is the first paper to revisit the topic with an eye on the modern mouse.
Besides informing manufacturers and end-users about the matter, we see two other, more scientific, motivations.
Firstly, it is necessary to return to empirical research into the effect of sensor position with a study that is adequately reported upon and adheres to today's methodological standards.
Secondly, 
the optical mouse allows designing a device that can \emph{change} the position of the sensor from the software side.
We present constructions for such a device, which we exploited for systematically finding the most suitable position for an individual user.

The paper is organized as follows. After discussing relevant literature, we
introduce the construction of two variable-sensor-position mice, one based on a sliding structure for the mechanical part, and another using a virtual sensor model with two sensors that uses a signal fusion technique.  
The virtual sensor allows changing the sensor position via software.
Using our system, we estimate the effect of sensor position on user performance in pointing. 
Finally, we offer two calibration methods for sensor position optimization. 
The two takeaways for practitioners are 
1) that placing the sensor in the center is the best compromise for regular users and
2) that a variable-sensor-position mouse can further improve the pointing throughput from that with a fixed center position.

%% file: sec_relwork.tex
The engineering and ergonomics of the computer mice have been a topic of HCI research for decades.
We consider three core device-related factors here: shape, weight, and transfer function.
Reliable results have been obtained that link changes in these factors to ergonomics and movement performance.
Additionally, we review previous attempts of integrating two sensors on a mouse, which we employed to implement the variable-sensor-position mouse.

\subsection{Effect of shape}
Shape has been the most intensively studied factor. 
Continuous use of a mouse, especially in dragging, is known to increase carpal tunnel pressure \cite{keir1999effects} and potentially lead to carpal tunnel syndrome. 
Slanted or vertical designs, also known as an ergonomic mouse,
are known to reduce pronation of the wrist and muscle use
 but produce adverse effects on pointing performance in general \cite{Chen2007ey, odell2007evaluation, gustafsson2003computer, quemelo2013biomechanics}. 
Some research has shown that a vertical (``ergonomic'') shape can indeed be beneficial ergonomically \cite{Hedge2010ep}, 
and performance approaches the levels of a regular mouse design with practice \cite{houwink2009providing}. 
Other attempts, such as a pen-like design \cite{dehghan2013designing} or alternative button designs \cite{lee2007alternative}, have been tried in efforts to minimize fatigue levels with mouse use.
Besides the slant angle, Isokoski and Raisamo \cite{isokoski2002speed} studied the effect of mouse size, interface (USB or PS/2), and sensor technology (ball or optical).
They found no practical performance differences among these. 
However, it was later discovered that the size of a regular mouse is not suitable for children \cite{Hughes2012}.
Hand size must be considered in choosing a comfortable mouse. 
Shape-changing mouse designs have been proposed also, such as the \textit{Inflatable} mouse \cite{kim2008inflatable} and the \textit{Adaptive} mouse \cite{tang2010adaptive}.
While we look at sensor position, not shape, we share the goal of adapting this factor to the user.

\subsection{Effect of weight}

A mouse weighs from about 52 g (e.g., Cooler Master MM710) to over 150 g (e.g., Logitech G602 with battery).
This is another important specification for a choice of mouse and, similarly, interacts with biomechanical considerations. 
Some models provide extra ballast so that a user can tune the weight distribution.
However, we could find only a few academic studies on this topic. Chen et al. \cite{chen2012weight} found that mouse weight affects the range of motion (ROM) of the wrist in a rapid targeting task.
A V-shaped relationship, centered at 130 g, was found between weight and ROM. 
Excessively light or heavy mice tend to induce ulnar-side shifts over time, which results in rapid radial-side corrective motion. 
Cabe\c{c}as \cite{cabeccas2010friction} found that a heavy mouse may increase friction between the mouse and the pad, thereby necessitating more muscle effort.
Still, no rigorous performance study of the effect of weight on performance has been conducted, 
so the relationship between weight and performance remains unresolved.

\subsection{Effect of transfer function}
The CD gain function, also known as the transfer function, maps between the mouse motion and the cursor movement.
The goal has been to find the ``best'' transfer function -- i.e., one that maximizes throughput \cite{fitts1954information} in pointing.
When a constant gain function is used, the ratio between the resolution of the mouse and screen has a single gain value. 
An appropriate gain value is a level that does not cause clutching (or rely on imprecise limbs for control) and involves the correct quantization approach \cite{casiez08}. 
If the gain is set too low, the mouse cursor moves so slowly that a one-stroke mouse movement is not enough to reach the target; therefore, clutching arises. 
Limb precision imposes a motor control limit, which is estimated to be around 700--1400 units per inch in the case of a mouse  \cite{berard2011humanLimit}. 
Quantization problems occur when the sensor resolution is not high enough to address one unit on the display. 
Commonly, modern mice are in the 800--1000 CPI range, and display resolution is 72--100 PPI (pixels per inch), though some high-density displays (e.g., the Apple Retina Display) may reach 300+ PPI. 
A proper gain should take both the device and display resolution into consideration.
Non-constant gain functions, often called \textit{pointer acceleration}, have been investigated for solving the clutching and the quantization problem at the same time. 
The general idea is to increase the CD gain in accordance with the speed of motion. 
Thus, a cursor can \textit{jump} toward a distant target with high-speed motion and address a small target at low speed. Each operating system (MacOS, Windows, Linux, etc.) provides a uniquely shaped gain function \cite{casiez2011no}.
Beyond those default presets, a recent work named AutoGain \cite{lee2020autogain} introduced an adaptive method for personalizing gain function.

\subsection{Mouse with two sensors}
Having two sensors on a mouse allows capturing the final missing piece in 2D motion: rotation. 
As stated before, conventional mouse sensors ignore rotation: the ball-and-wheel mechanism cannot capture it by principle, and the optical sensors ignore it by design (in fact, detecting rotation on a single optical sensor is possible \cite{poston2007computer}). 
MacKenzie et al. proposed the earliest prototype mouse with two balls in 1997 \cite{MacKenzie1997} to support three degree-of-freedom (3DoF) interaction, e.g., translate and rotate an object at the same time. A series of studies followed \cite{Apperley2013, almeida2006supporting, fallman20073dof, hannagan2007twistmouse, Lee2015, poston2007computer}. However, 3DoF manipulation was better supported by a wheel \cite{Apperley2013}. There is evidence that a major mouse company, Logitech, tested a two-sensor mouse for 3DoF support \cite{logitech2008}, but no mouse was successfully commercialized so far \cite{mackenzie2015user}. 
One inspiring work from Lee and Bang \cite{Lee2015} exploited the two sensors to compensate for coordinate disturbance from the kinematic rotation of the human arm. The goal of their work was not to introduce a new DoF, but to support better performance in a regular drawing task.
In this paper, the purpose of two-sensor implementation is only to simulate a physical sensor on an arbitrary location. Similar simulation models were introduced previously \cite{Apperley2013, poston2007computer}, but they have not been verified. We validate the model theoretically and empirically.
The virtual sensor implementation allows better mechanical stability from no moving part, and opens up a new possibility of the software-side optimization procedure.

\subsection{Summary of prior work}  
While, to the best of our knowledge, the effect of sensor position has not been studied after the original study by Verplank and Oliver on the Microsoft Mouse \cite{verplank1989microsoft},
other findings -- on the effects of joint rotation -- hint at the possibility of an empirical effect.
Studies of computer ergonomics have revealed that the rotation of a mouse is connected with rotation in wrist and arm posture. 
Large radial and ulnar deviations in the elbow and the wrist are common,
with wrist deviation ranging from 10 to 25 degrees \cite{burgess1999wrist, wahlstrom2000differences, jensen1998job}. 
Those values are large, given that the range of motion in radial--ulnar deviation is about 30 to 40 degrees \cite{chaparro2000range, chen2012weight}. 
In the extreme case, if accompanied by elbow rotation, 
a mouse could rotate by up to 60 degrees during common usage.
These observations about rotation provide an incentive for studying the effect of sensor position.

%% file: sec_virtual_sensor.tex
\begin{figure}[tb!]
	\centering
	\includegraphics[width=.999\columnwidth]{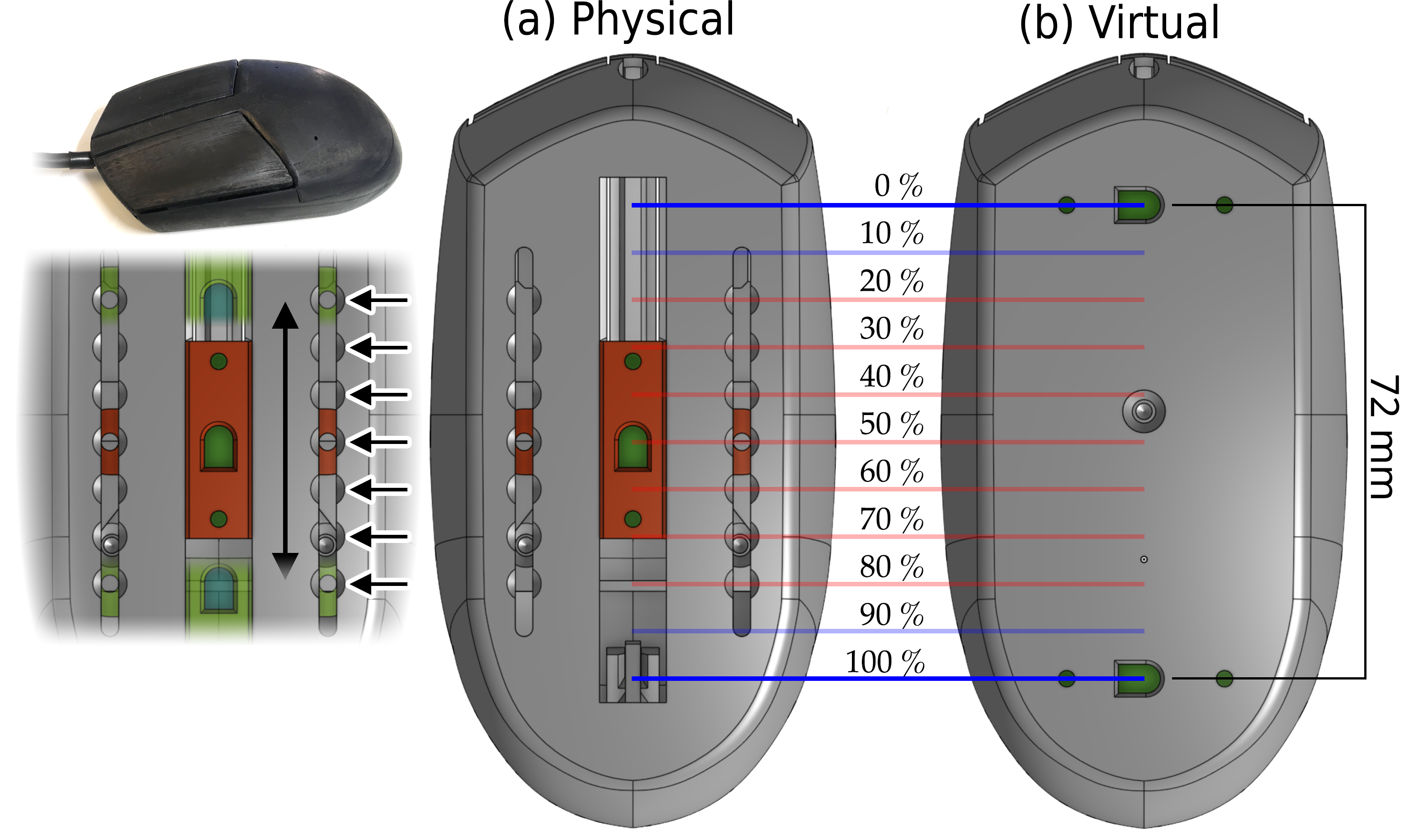}
	\caption{Constructions of variable-sensor-position mice. (a) \textit{Physical}, in which the displacement sensor is on a sliding rail. (b) \textit{Virtual}, with two sensors to compute a virtual sensor at any location in between. We label the extreme front position as 0\% and the rearmost position as 100\%. In the \textit{Physical} device, the motion sensor can be placed between 20\% and 80\% position, in 10\% increments.}~\label{fig:mouse_devices}
	\vspace{-2em}
\end{figure}

We demonstrate two constructions of \emph{variable-sensor-position} mice.
An overview is given in Figure \ref{fig:mouse_devices}.
The physical sensor (Approach 1) allows manually sliding the sensor along a rail, in similarity to the \textit{Ninox Astrum} in Figure \ref{fig:sensor_positions} but with a wider range. 
The virtual sensor (Approach 2) is able to emulate any sensor position on the vertical axis, and the position of the sensor can be changed in software. 
Below, we present the principles of emulating a sensor and our empirical testing of the accuracy achieved.

\subsection{Approach 1: Physical sensor positioning}

\emph{Mouse device:} The dimensions of the shell are $117 \times 62 \times 38$ mm
while the shape matches that of a popular commercial mouse (Logitech G102/G203; our implementation does not have a scroll wheel). 
Our shells were 3D printed, including buttons and mounts for switches and sensors.
We designed the sensor mounts to align the optical center of the sensors with the centerline of the device.
We used a PixArt PMW3360 optical 2D displacement sensor, a model frequently used in high-end gaming mice. 
The sensor reads incremental horizontal ($dX$) and vertical ($dY$) displacement.

We used two Omron D2F-F microswitches, mounted under the left and right buttons.
A SparkFun Pro Micro board interfaced with the sensor module via SPI protocol.
The board sampled the sensor data at 500 Hz. 
The sensor resolution was set to the maximum value (12,000 CPI) internally. 
Users can select their preferred resolution (\texttt{UserCPI}), and the raw readings from the sensor are down-scaled to that value for USB reports.
The microprocessor (Microchip ATmega32u4) supports native USB communication. Two USB endpoints were implemented: an HID (Human Interface Device) mouse, and a CDC (Communication Device Class) serial for logging. 

\emph{Sliding rail:}
The sensor module was attached to a rail at the bottom of the shell.
The rail offers seven mounting positions at 7.2~mm intervals between 20\% and 80\% positions (marked by arrows in the figure). 
The sensor can be manually moved.

\emph{Tracking:}
In each sample, a timestamp, two button states, two raw sensor values ($dX$, $dY$), and their down-scaled values ($mX = k \times dX$, $mY = k \times dY$) are logged through CDC serial, where $k$ is a CPI multiplier calculated as $k=(\texttt{UserCPI} / 12,000)$.
Separately, an HID report packet consists of the down-scaled values ($mX$, $mY$) and the two-button states, which are sent to the host computer and controlled the system cursor. 
Because it only allows integer values, truncation occurs at the decimal point, and the remainder is added to the next sample.

\subsection{Approach 2: A virtual sensor}
All properties of the \textit{Virtual} device (Figure \ref{fig:mouse_devices}, right) are identical to those in Approach 1 except that
there are two fixed sensors instead of one sensor on a sliding rail. 
The front and the rear sensor measure displacement separately, 
denoted as ($dX_{front}$, $dY_{front}$) and ($dX_{rear}$, $dY_{rear}$), which allows emulating a virtual sensor position ($mX$, $mY$) by means of Equation \ref{eq:virtual_mouse}:
\begin{equation}
 \label{eq:virtual_mouse}
\begin{aligned}
mX &= k\{(1 - p) dX_{front} + p \cdot dX_{rear}\}  \\
mY &= k\frac{dY_{front} + dY_{rear}}{2}
\end{aligned}
\end{equation}
The $p$ value, denoting the emulated position of the sensor, can be set within the range 0--100\% with 1\% resolution.
In each sample, a timestamp, two button states, four raw sensor values ($dX_{front}$, $dY_{front}$, $dX_{rear}$, $dY_{rear}$), and the down-scaled virtual sensor values ($mX$, $mY$) are logged. 
An HID mouse report is generated as in Approach 1. 
In following, we show Equation~\ref{eq:virtual_mouse} forms a virtual sensor equivalent to the physical sensor.

\vspace{-.5 em}
\input{sec_virtual_sensor_proof}

\subsection{Accuracy measurements with a robot}
The virtual sensor model assumes ideal sensors that only measure translational movement.
However, a real sensor may react to rotational movement as well.
We wanted to test the virtual sensor model empirically, using an instrumented mouse.
We assessed the accuracy of the \textit{Virtual} device against a ground-truth device (\textit{Physical}) by comparing trajectories produced by a high-precision robotic arm.

\textbf{Method:} The two mice were modified to have DIN ISO 9409-1-A50 mounts on top of their cover. 
The center of the mount, the axis of rotation, was placed at the $p=50\%$ position.
Each device was attached to a FRANKA EMIKA\footnote{\url{https://www.franka.de}} Panda robotic arm, which has a reported accuracy of $\pm$.1~mm.
It was set to move in $\infty$ shape with and without rotation. 
See the top row in Figure \ref{fig:physical_vs_virtual} for the two programmed motions employed. Planned path lengths of the both were 700 mm.
When rotation was applied, the mouse rotated by -20$\degree$ at the leftmost position and +40$\degree$ at the rightmost position.
As the robotic arm moved the device, a host computer collected raw sensor logs: 
($dX$, $dY$) data for the seven mounting positions (20\%--80\%) with the \textit{Physical} device; 
($dX_{front}$, $dY_{front}$) and ($dX_{rear}$, $dY_{rear}$) data from the 0\% and 100\% positions with \textit{Virtual}. 
We simulated the cursor movements for the 20\%--80\% positions from the \textit{Virtual} device by using Equation \ref{eq:virtual_mouse} with $k=1$.
We repeated the execution three times.
The repeated measurements produced virtually identical data, so we picked only the last measurement for further analysis. 

\textbf{Results:} 
We compared \textit{Virtual} and \textit{Physical} in all sensor-position conditions between 20\%--80\%. See Figure \ref{fig:physical_vs_virtual} and Table \ref{tab:robot_exp_result} for the results.
We note that the data from \textit{Physical} exhibited slightly more deviation, because of an angular and positional fixture error at each mounting position.
Looking at the translation-only motion (see Figure \ref{fig:physical_vs_virtual}, left column) revealed trajectories that overlap almost perfectly. The measured average trajectory length was 709 mm, which was only marginally different from the planned length of 700 mm.
When rotation is added (see Figure \ref{fig:physical_vs_virtual}, right column), the cursor movements are distorted \cite{Lee2015}. 
Moreover, changing the sensor placement altered the cursor trajectory systematically. 
As the sensor moved from the front to the rear (toward 100\%), 
the cursor moved less in the \texttt{x} direction. As a result, the sensor at 20\% traveled 6.9\% longer compared to the sensor at 80\%.
We can also confirm the simulated cursor motions overlapped the real one with tiny discrepancy (<1\%).

\afterpage {
\begin{figure}[t!]
	\centering
	\includegraphics[width=.99\columnwidth]{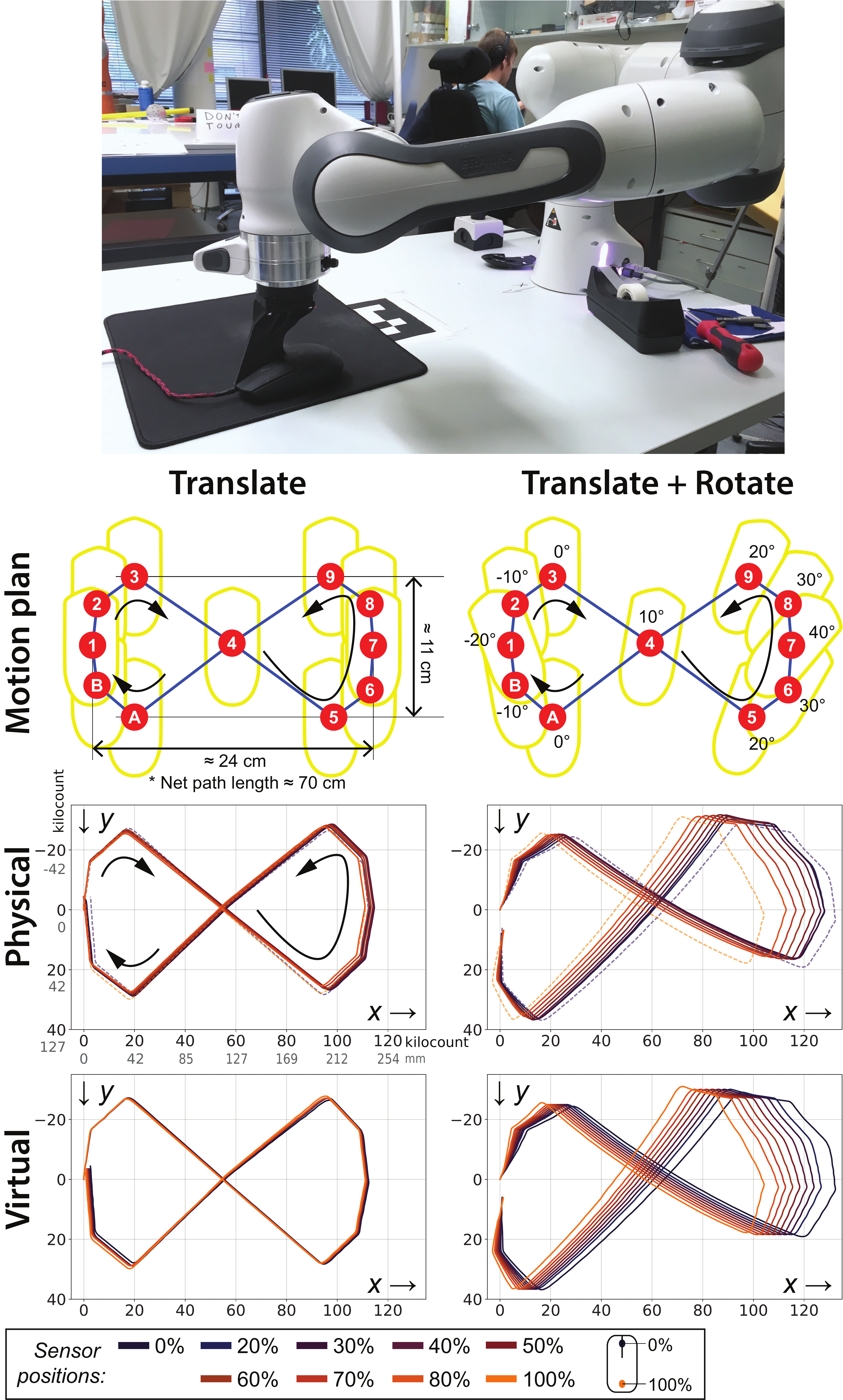}
	\caption{The virtual-sensor method was assessed by comparing trajectories against a corresponding physical-sensor mouse. A robotic arm (top) performed $\infty$-shaped trajectories (top row, 1--2--3--4--5--6--7--8--9--4--A--B--1) without (left column) and with rotation (right column). 
		The virtual sensor (bottom row) generated nearly identical cursor trajectories relative to the physical sensor (middle row). Note: Units are kilocount, where 1 kilocount $\approx$ 2.12~mm with the 12,000 CPI resolution sensor.}~\label{fig:physical_vs_virtual}
\vspace{-1em}
\end{figure}

\begin{table}[t!]
	\centering
	\begin{tabular}{l | r r | r r}
		& \multicolumn{2}{c |}{\small{\textbf{Translate only}}} & \multicolumn{2}{c}{\small\textbf{Translate+Rotate}}\\
		& {\small \textit{Physical}}
		& {\small \textit{Virtual}}
		& {\small \textit{Physical}} 
		& {\small \textit{Virtual}}\\
		\midrule
		Avg. length	\small{(kilocount)}	& 335.1	& 335.0	& 341.7	& 338.5
 \\
		\small{~- Standard Deviation}	& \small{3.76} & \small{.14} & \small{8.45} & \small{7.94} \\		
		\small{~- in mm @ 12k CPI}		& \small{709.4} & \small{709.1} & \small{723.4}	& \small{716.5}	 \\
		Length @ 20\% position 			& 333.1	& 335.2	& 347.2	& 35.7  \\
		Length @ 80\% position			& 328.0	& 334.8	& 326.0	& 326.9  \\
		\begin{tabular}{@{}l@{}}Mean path discrepancy \\ ~\small{ of \textit{Virtual} from \textit{Physical}} \\ \end{tabular}			
				& \multicolumn{2}{r |}{1.33 \small{(= .40\%$^{*}$)}}
				& \multicolumn{2}{r}{3.12 \small{(= .91\%$^{*}$)}} \\
		Avg. motion duration			& \multicolumn{2}{c}{9.2~s} & \multicolumn{2}{c}{11.8~s} \\
		\multicolumn{5}{r}{\small{Unit: kilocount, if unspecified / *Compared to the path length}} \\
	\end{tabular}
	\caption{Results of the accuracy measurement experiment. The values are calculated from the detected movements from the sensors.}~\label{tab:robot_exp_result}
\end{table}
}
 
\subsection{Human-subject observation}
Even though the robot experiment exhibited an effect of device rotation on cursor trajectory, we wanted to also check the effect with humans subjects.
We conducted a study of mouse usage pattern while playing a simple game.

\begin{figure}[tb!]
	\centering
	\includegraphics[width=.8\columnwidth]{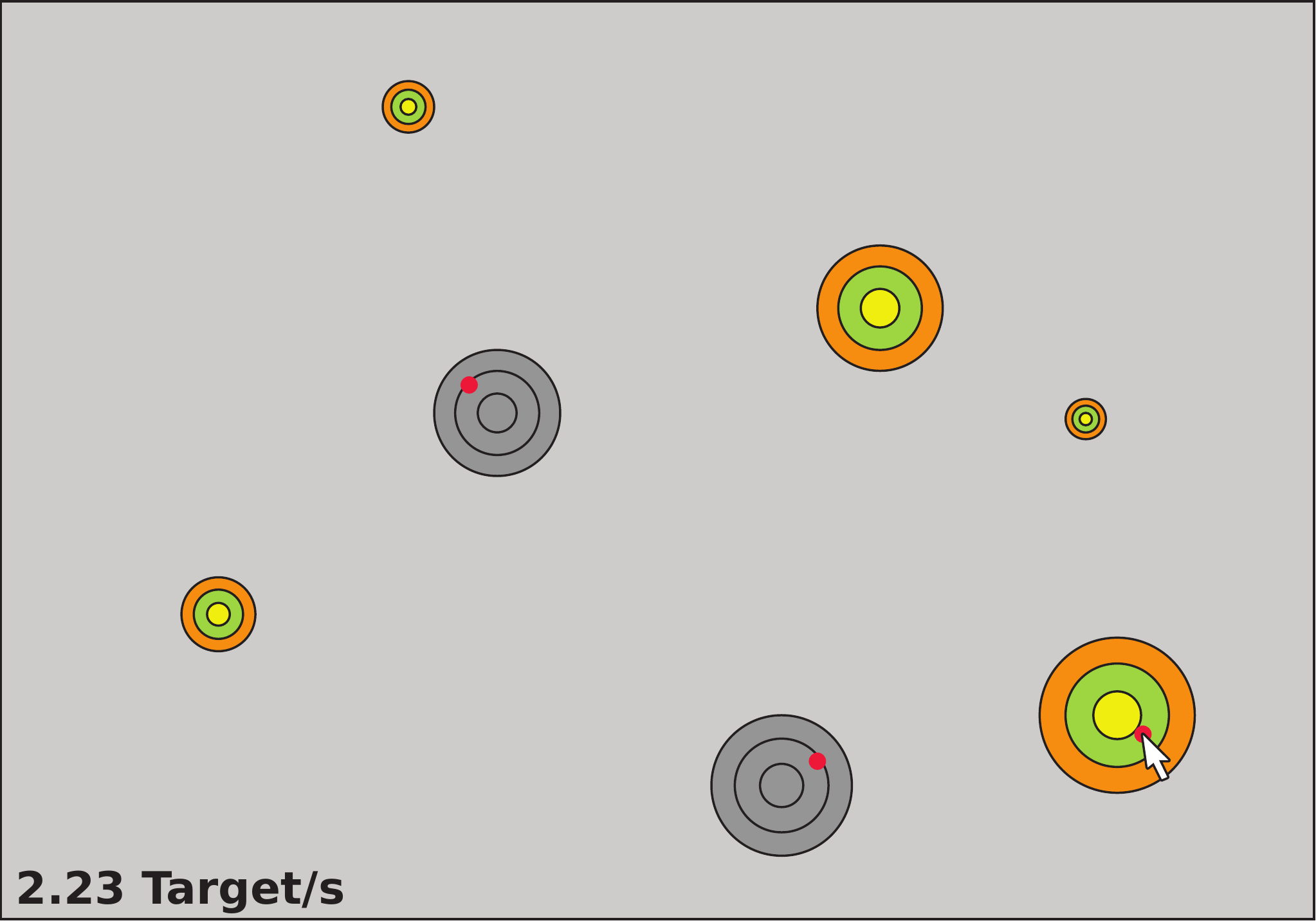}
	\caption{The aimed-movement game AimBooster (in \textit{Auto-balanced} mode with the default targets/second setting, \url{http://aimbooster.com}) was used to assess the sensor models empirically. The goal of the game is to click targets that spawn in random positions.}~\label{fig:aimbooster}
	\vspace{-1em}
\end{figure}

\textbf{Method:} 
Six participants (age Mean 30.2, \emph{SD} 5.9; 5 male, 1 female) played the game shown in Figure \ref{fig:aimbooster}, using the virtual sensor mouse.  
The virtual sensor was fixed to 50\% position (set $p=0.5$ in Equation \ref{eq:virtual_mouse}).
In the game, targets are spawned in random positions, 
each persisting for 3.5 seconds. 
Users were requested to click all the targets before they disappear. 
The initial spawning rate was set to 2 targets/s, 
and the speed gradually increased or decreased in response to the performance, targeting 92\% success rate.
During a five-minute practice session, 
each participant was asked to set the sensor resolution to his or her preferred value. 
Three-minute test session was followed and samples were logged. 
We asked participants to play as quickly and accurately as possible. 
The average spawning speed reached by the participants in that session ($\approx$ click rate) ranged from 1.2 to 2.2 targets/s.

\begin{table}[b!]
	\centering
	\begin{tabular}{l | r r r | r r r}
		
		Participant 
		    & \multicolumn{3}{c|}{\textbf{$dX$}} 
		    & \multicolumn{3}{c}{\textbf{$dY$}} \\
		\cmidrule(r){2-4} \cmidrule(r){5-7}
		\small{@CPI}
		 & \small{Slope} & \small{Intercept} & \small{$R^2$} &
		 \small{Slope} & \small{Intercept} & \small{$R^2$} \\
		\midrule
		P1 \small{@1200}	& .55 & -.009 & .595 & 1.02 & -.006 & .989 \\
		P2 \small{@800}	    & .60 &  .073 & .845 & 1.01 &  .004 & .991 \\
		P3 \small{@1200}	& .52 & -.006 & .834 & 1.01 & -.008 & .990 \\
		P4 \small{@600}	    & .54 & -.008 & .932 & 1.01 & -.011 & .992 \\
		P5 \small{@1200}	& .55 &  .093 & .747 & 1.01 &  .003 & .989 \\
		P6 \small{@1200}	& .58 & -.165 & .776 & 1.00 &  .003 & .990 \\	
		\midrule
		Average & .56 & -.004 & .790 & 1.01 & -.003 & .990 \\
	\end{tabular}
	\caption{Linear regression analysis between ($dX_{front}$, $dX_{rear}$) and ($dY_{front}$, $dY_{rear}$) pairs for each participant. <1 slopes in $dX$ mean that the front sensor reads larger values than the rear sensor values, while the $dY$ values are almost identical (slope $\approx$ 1.0, $R^2$=.99). The intercepts are near zero (unit: count), meaning in-place rotation almost never happened.}
	\label{tab:dx_and_dy}
\end{table}

\textbf{Results:} Linear regression analyses were performed between $dX_{front}$--$dX_{rear}$ and $dY_{front}$--$dY_{rear}$ pairs. 
Table \ref{tab:dx_and_dy} shows that the choice of sensor position has an apparent effect on the displacement sensed. 
All participants were very similar in the patterns exhibited, whatever their choice of CPI. 
The rear sensor reads almost half the horizontal displacement ($dX$) read by the front sensor.  
For vertical displacement ($dY$), the sensors' readings were nearly identical. Low $R^2$ in $dX$ indicates users rotated the mouse inconsistently during gameplay, which implies the complex nature of the sensor position problem.

\begin{figure}[tb!]
	\centering
	\includegraphics[width=.8\columnwidth]{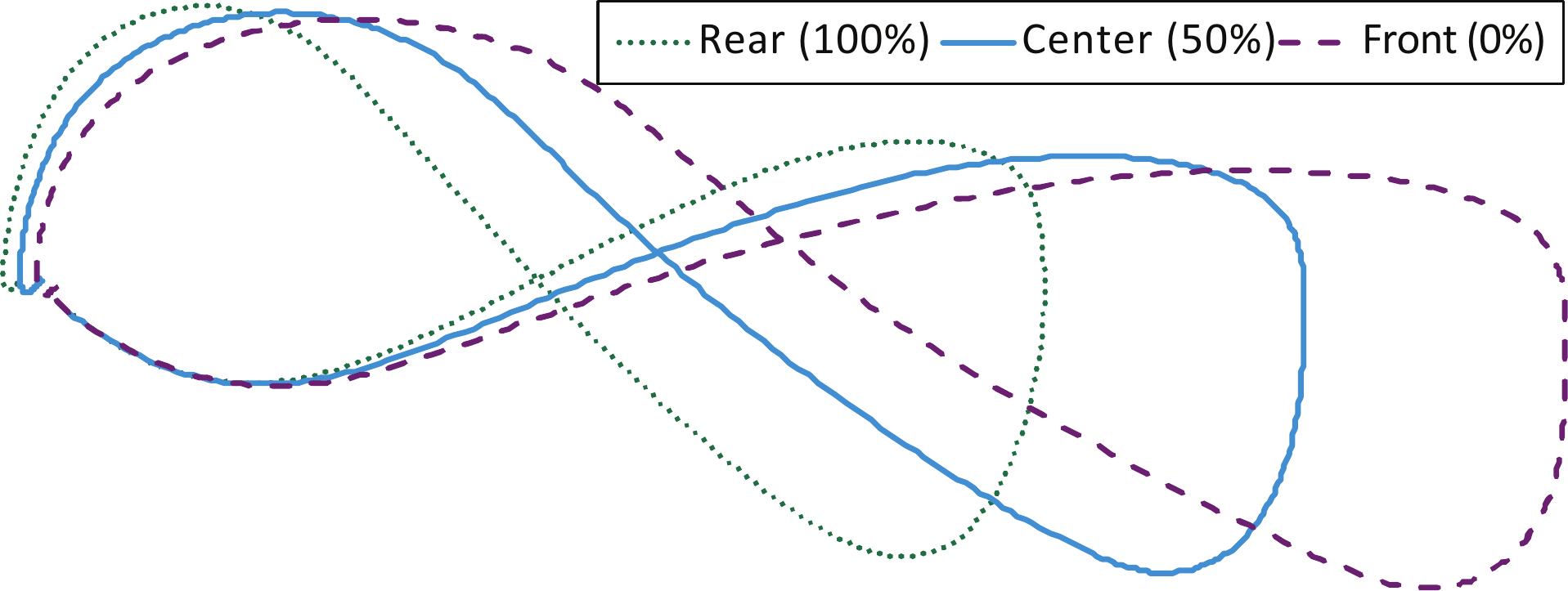}
	\caption{Hand-drawn $\infty$ shapes on different sensor positions. One subject (P4) drew the shape once. The front and rear data are raw data, and the center data is virtually synthesized using Equation \ref{eq:virtual_mouse}.}~\label{fig:drawing}  
\vspace{-2em}	
\end{figure}

With human subjects, the front sensor read much larger $dX$, up to 192\% of the rear sensor. For example, an $\infty$ shape was hand-drawn with a single motion (see Figure \ref{fig:drawing}), and the front sensor drew about twice wide shape compared to the rear sensor. In the robot experiment, this expansion was less pronounced (see Figure \ref{fig:physical_vs_virtual}). These results indicate the effect of sensor position is a practical problem, even beyond the assumption we made in the robot experiment.
The issue is more apparent when performing a directional movement: e.g., when a user aims a target at 45\degree~direction, the cursor may move towards a lower or higher angle depending on different sensor placements. See Figure \ref{fig:problem} for an illustration of this situation.

%

%% file: sec_virtual_sensor_proof.tex
\begin{proof}
Virtual and physical sensors at position $p$ yield identical measurements.

\begin{figure}[tb]
\centering
  \includegraphics[width=0.999\columnwidth]{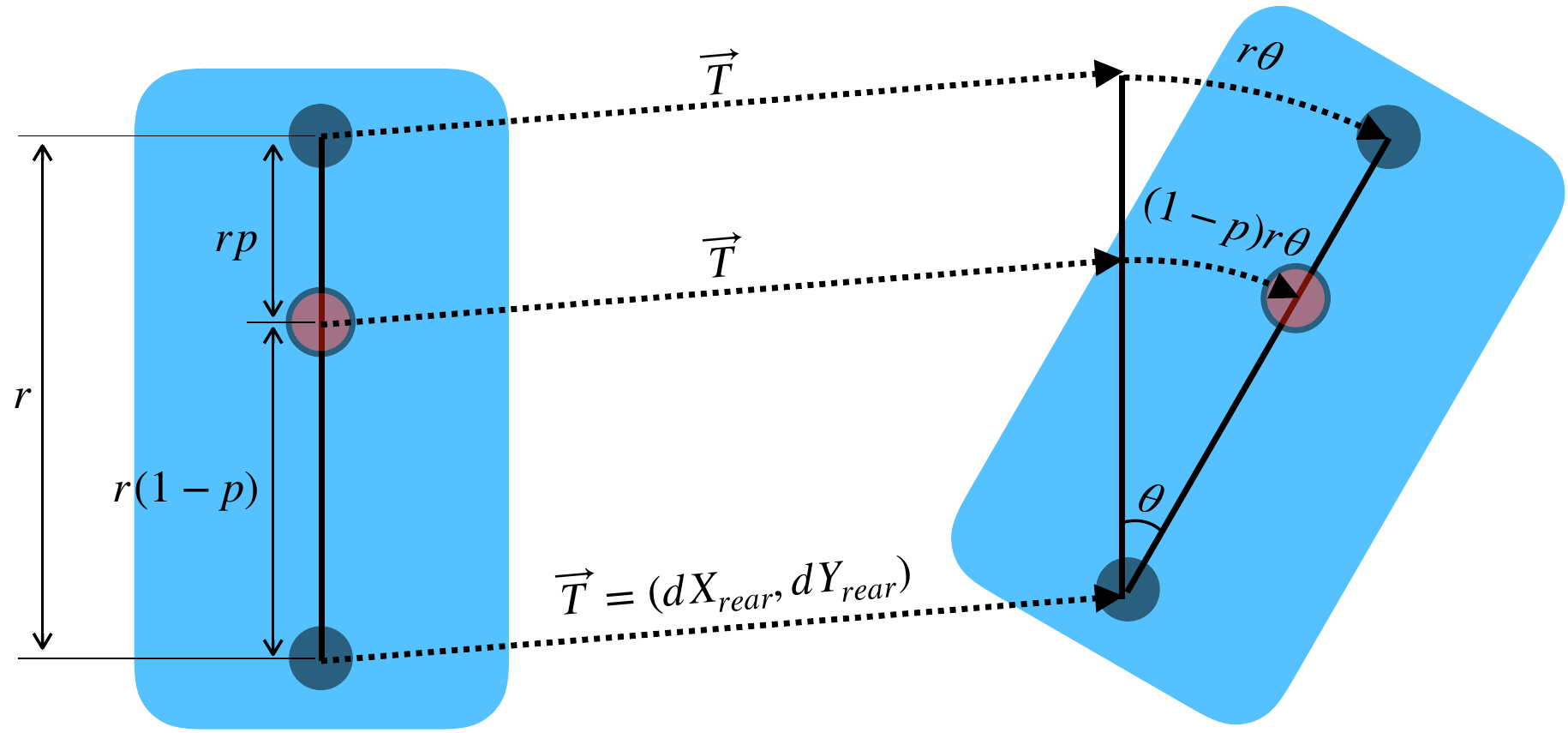}
  \caption{Illustration of a virtual sensor during motion. We assume physical displacement sensors (in gray) in the front and rear part of the device. A simulated virtual sensor (the red, black-outlined dot) is located between the two sensors. 
  	$r$ is the distance between physical sensors. 
  	$p$ is the position of the virtual sensor range between 0--100\%.
  	See the text for the calculations.}~\label{fig:virtual_proof}
  \vspace{-2em}
\end{figure}

\emph{Intuition:} Let us look at what happens between two samples in a moving device. 
Figure \ref{fig:virtual_proof} shows device locations in $(t-1)^{th}$ and $t^{th}$ samples. 
Let us set a reference point at the rear sensor. 
Now, its movement can be broken down into a translation vector $\vec{T}$ and a rotation of $\theta$ (unit: rad). 
With only translation movements considered, the same displacement vector $\vec{T}$ is read from both sensors. Next, during the rotation process, the rear sensor does not undergo any translation movement, so it does not measure any displacement.
However, the front sensor is translated further by the arc length ($r\theta$) of the rotation around the rear sensor.
Because the added translation is tangential with respect to the front sensor, the $r\theta$ value is added to the horizontal displacement for only the front sensor reading. 
Similarly, the imaginary virtual sensor first translates by $\vec{T}$, and the rotation adds $(1-p)r\theta$ to its measurement of horizontal displacement. 

The sensor readings are modeled by Equation \ref{eq:sensor_readings}. Amount of rotation $\theta$ and the simulated virtual sensor data can be derived from the two sensors as in Equation \ref{eq:sensor_theta} and Equation \ref{eq:sensor_virtual}.

\begin{equation}
 \label{eq:sensor_readings}
\begin{aligned}
\texttt{Rear} &= (dX_{rear}, dY_{rear}) = \vec{T}\\
\texttt{Front} &=  (dX_{front}, dY_{front})= (dX_{rear} + r\theta, dY_{rear}) \\
\texttt{Virtual} &= (dX_{rear} + (1-p)r\theta, dY_{rear})
\end{aligned}
\end{equation}
\begin{equation}
 \label{eq:sensor_theta}
\begin{aligned}
dX_{front} - dX_{rear} &= (dX_{rear}+r\theta) - (dX_{rear})= r\theta \\
               \therefore\;\; \theta &= \frac{dX_{front} - dX_{rear}}{r}
\end{aligned}
\end{equation}
\begin{equation}
 \label{eq:sensor_virtual}
\begin{aligned}
\texttt{Virtual} &= (dX_{rear} + (1-p)r\theta, dY_{rear}) \\
        &= (dX_{rear} + (1-p)r\frac{dX_{front} - dX_{rear}}{r}, dY_{rear}) \\
        &= ((1-p)dX_{front} + p \cdot dX_{rear}, dY_{rear})
\end{aligned}
\end{equation}

In addition to Equation \ref{eq:sensor_virtual}, we replace $dY_{rear}$ with $\frac{dY_{front}+dY_{rear}}{2}$ although they are expected to be same. By averaging them, we compensate a possible deviation between sensors. Finally, we multiply a CPI multiplier $k$ to both the $X$ and $Y$ values for down-sampling, which yields Equation \ref{eq:virtual_mouse}. 
\end{proof}

%% file: sec_experiment.tex
We wanted to rigorously measure the effect of sensor position in a controlled pointing task.
The previously published study, comparing three sensor positions, with three different mouse shapes, was from 1989, and five of the 12 participants had never used a mouse \cite{verplank1989microsoft}. 
Our experiment adheres closely to the ISO 9241-411:2012 (previously ISO 9241-9:2000) standard for evaluation of input devices
with minor differences: (1) the calculation of the effective index of difficulty $ID_e$, and (2) the treatment of outliers \cite{wobbrock2011effects}.
The test is used to evaluate pointing performance in different movement directions.
To understand the systematic effect of sensor position,
we varied it in the full range of available positions in 20\% increments (=14.4 mm), plus the center (50\%) which is the most common choice for commercial mouse devices (see Figure \ref{fig:mouse_devices}).

\subsection{Method}
We used a within-subject design. Sensor position is an independent variable, having seven levels: 0\%, 20\%, 40\%, 50\%, 60\%, 80\%, and 100\%. Pointing throughput and path deviation are dependant variables.

\textbf{Participants:}
We recruited hobby gamers, who represent users who are potentially interested in optimizing mouse performance.
We recruited 14 participants who regularly play games using a mouse (age Mean 27.8, \emph{SD} 5.6; nine males and five females). 
All participants had normal or corrected-to-normal vision and were right-handed. 
They used a mouse daily: 1.6 h (\emph{SD} 0.8) for gaming and 3.5 h (\emph{SD} 1.8) for general tasks. 
Their hand sizes (measured from the tip of the middle finger to the midcarpal joint) ranged from 15.5 cm to 20 cm (Mean 17.3, \emph{SD} 1.4). 
Their regular mice had sensors around the center, with some variations around 40\%--60\% positions. No devices featured an extreme sensor design, such as in the Apple Magic Mouse. 
All participants signed an informed consent form.
Each was given a movie ticket equivalent to 14 EUR for participating.

\textbf{Apparatus:}
The virtual-sensor mouse was used. \texttt{UserCPI} was fixed at 800, a common value used in practice. A constant gain function was used, which translates 1 count = 1 pixel.
Besides adding realism, this decreases the possibility of our results being biased from the choice of transfer functions.
An experiment program was implemented in \texttt{Processing}\footnote{\url{https://processing.org/}} and ran in full-screen mode.  
We used a high-speed gaming monitor (BenQ XL2546, 24-inch, 1920$\times$1080~px, 92 pixels~per~inch, 240 frames~per~second).  
A desktop computer (Intel Core i7 8700 3.2~GHz, 32~GB RAM, NVIDIA GeForce RTX 2080) drove the experiment program.  

\begin{figure}[tb!]
	\centering
	\includegraphics[width=0.99\columnwidth]{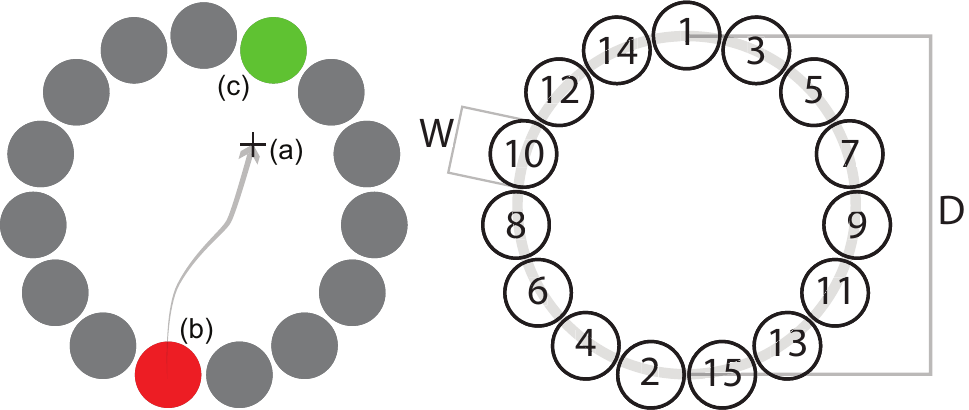}
	\caption{The multi-directional tapping test was used to measure the effect of sensor position. 
		\textbf{Left}: Each trial consists of cursor movement (a) from the previous target (b) to the current target (c) and a button click. The current target is highlighted in green. If the previous attempt yielded a miss, the previous target turns red. \textbf{Right}: Definitions of $D$ and $W$, and the order of the targets.
	}~\label{fig:fitts}
	\vspace{-1em}
\end{figure}

\textbf{Task:}
We adopted the multi-direction tapping test from ISO 9241-411:2012, with 15 round targets in a circular arrangement (see Figure \ref{fig:fitts}). 
Distance ($D$) is defined as the diameter of the outer circle, and target width ($W$) is defined as the diameter of each target. 

At the beginning, the first target, \textcircled{\small{1}} (see Figure \ref{fig:fitts}, Right), is highlighted in green. Clicking on it starts a \textit{session}, and immediately the next target, \textcircled{\small{2}}, gets highlighted. The participant moves the cursor (\textcircled{\small{1}}$\rightarrow$\textcircled{\small{2}}) and clicks it. This is defined as one \textit{trial}. If the click is outside the target (a mistake), it is marked in red and no correction is allowed. After all the targets are visited,  clicking the first target again (\textcircled{\small{15}}$\rightarrow$\textcircled{\small{1}}) completes the session. Accordingly, one session strictly consists of 15 trials and includes 16 mouse clicks.

We used $D$ = \{300, 900 px\} $\times$ $W$ = \{20, 50, 120 px\}\footnote{{Physical sizes: $D$=\{83, 248 mm\}, $W$=\{5.5, 14, 33 mm\}}} =~6~combinations, designed to distribute the index of difficulty ($ID$) over the range 1.8--5.5 bits. Each combination was repeated three times; therefore, there were $6\times3=18$ sessions in one \textit{Position block}.
In total, 7 (blocks) $\times$ 18 (sessions) $\times$ 15 (trials) = 1,890 trials were collected per participant. 

The order of the sensor positions (blocks) was counterbalanced across participants via a balanced Latin Square design \cite{williams1949balanced}.
Within each block, the $D$ and $W$ combinations were presented in random order.

\textbf{Procedure:}
After a participant arrived, an experimenter explained the purpose of the experiment: ``We are testing the pointing performance with different mouse settings.'' We did not mention that the mouse sensor position would be the setting changed.
The task was demonstrated, and participants practiced it with a default mouse setting (800 CPI, sensor position 50\%) for 10 sessions with random $D$ and $W$. During the practice, they adjusted the position of the desk, chair, mouse, and monitor to be comfortable.

The experimenter configured the sensor position before each block.
To adapt initially to the sensor configuration presented, participants played the AimBooster game (see Figure \ref{fig:aimbooster}) for at least three minutes.  
They were encouraged to strive for >2.5 targets/s (<400 ms between clicks, 500+ clicks in total). 
We enforced this training to ensure that users adapted to the new condition.
After the adaptation period, one block of the task was completed: 2 ($D$) $\times$ 3 ($W$) $\times$ 3 (instances) = 18 sessions, in random order. 
The participants could rest whenever they wished between sessions. 
If the success rate within a block fell below 90\%, the block had to be redone.
A one-minute rest period was enforced between blocks.

Finally, the experimenter collected participants' demographics and took pictures of their dominant hand holding the mouse. 
The full procedure took about one hour. 

\textbf{Data preprocessing:}
Each trial produced a log entry with the previous target, the current target, and the timestamped cursor trajectory.  
In all, 26,460 trials were carried out.
We found and screened out 39 outliers, defined by these criteria: 
(1) the movement distance being less than $D/2$ \cite{wobbrock2011effects};   
(2) the start point or endpoint being more than $W\times2$ from the desired position \cite{wobbrock2011effects}; 
or (3) movement time exceeding $\log_2(D/W+1)$, which means an extremely slow trial with <1 bit/s throughput.

\subsection{Results}

The grand success rate was maintained at 94.7\% ($SD$=1.5\%, minimum 91.9\%) \cite{mackenzie1992fitts}.
For the following statistical analyses, 
the significance level $\alpha$ was set to .05 unless otherwise stated. 
The error bands in the graphs are 95\% confidence intervals (CI). 
Statistical tests were carried out with SPSS version 18 and Python SciPy module version 1.2.0.

\textbf{Throughput:}
Fitts' throughput (TP) is a performance metric for both speed and accuracy, calculated in terms of information transfer in the pointing task \cite{mackenzie1992fitts}. For post-hoc treatments for accuracy, bivariate endpoint deviation $SD_{x,y}$ \cite{wobbrock2011effects} was calculated for each session to estimate effective width $W_e = 4.133 \times SD_{x,y}$. Similarly, effective distance $D_e$ was calculated by averaging the movement distances within a session \cite{soukoreff2004towards}. 
 The effective index of difficulty of a session was calculated as $ID_e = \log_2(D_e/W_e+1)$ \cite{soukoreff2004towards}.

\begin{figure}[t!]
	\centering
	\includegraphics[width=1.0\columnwidth]{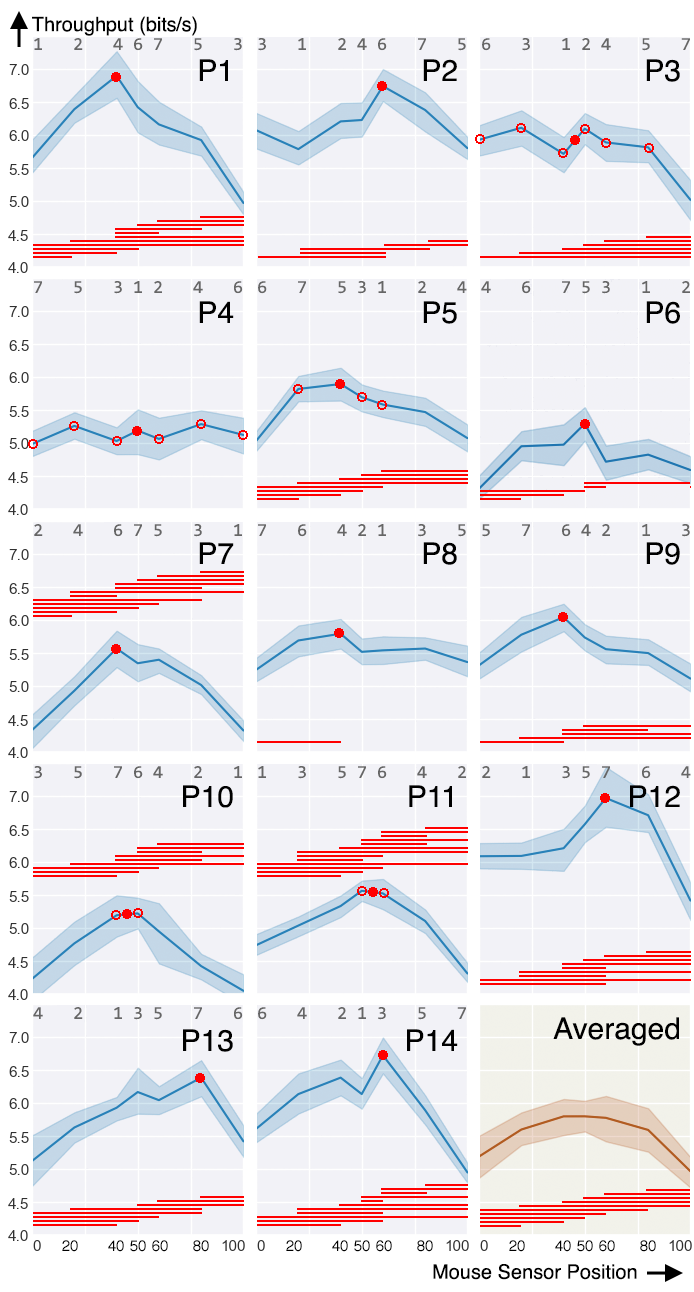}
	\caption{For most users in our study, the sensor position exhibited a significant effect. The P1--P14 graphs are TPs of each participant, and the \textbf{Averaged} graph shows the mean TPs averaged overall.
		The horizontal bars indicate significant differences. Error bands are 95\% CI.
		The solid red marks show the best sensor position found for each participant. The outlined marks are for positions in the best homogeneous subset if it has multiple items. The numbers at the top of each subplot are the order of blocks, which were counterbalanced across participants.} ~\label{fig:throughput}  
	\vspace{-2em}
\end{figure}

One data point ($ID_e$, $MT$) was produced from a session of 15 trials, where $MT$ is the averaged movement time. The throughput for a session $k$ was calculated as $\textit{TP}_k=ID_{e_k}/MT_k$, and for a block, we followed the mean-of-means approach to calculate $\textit{TP}=\frac{1}{N}\sum_{i=1}^{N}\textit{TP}_i$, where $N=18$ sessions in a block.

The observed TPs ranged from 4.0 to 7.0 bits/s (Mean 5.5, $SD$ .6). 
We performed a regression analysis for each block \cite{soukoreff2004towards}. The Fitts' law model was well fitted (Mean $R^2=.91$, $SD$ .04), with near-zero intercepts (Mean .016 s, $SD$ .057) as expected.

Figure \ref{fig:throughput} illustrates TPs of each participant (P1--P14) for the range of tested sensor positions.
We compared the positions within a participant by using Friedman's chi-square test, followed by a \emph{post-hoc} Tukey's HSD test. 
The Friedman test revealed a significant effect of sensor position on TP within each individual except P4 ($\chi^2=5.92$, $p=.43$). 
This means that every user except one showed a strong effect of the sensor position.
Using the \emph{post-hoc} comparison results, we calculated the best subset of conditions and picked a single best position for each individual (see the caption of Figure \ref{fig:throughput}).
The last plot (Averaged) is the mean TP over all the participants. 
Repeated measure analysis of variance (RM-ANOVA) revealed a significant effect of sensor position on TP ($F_{(3.51, 45.6)}=20.83$, $p<.001$, $\eta^2_p=.616$ with Greenhouse-Geisser correction). 
The TP values followed a quadratic distribution ($p<.001$).
Bonferroni \emph{post-hoc} analyses showed the extremal (0\% and 100\%) sensor positions to be worse than all others, but no significant difference was found in the positions between 20\% and 80\%. Compare to the 50\% position (TP=5.77), 100\% position was worse by 14.0\% (TP=4.97), and 0\% was worse by 10.7\% (TP=5.16).

We further analyzed the effect of sensor position for each individual. We performed two-way ANOVA on the TPs for all the sessions, grouped by \textit{Participant} and sensor \textit{Position}. 
Because the observations are not independent, we focused on effect size instead of on statistical significance (all $p$s <
.001). 
The partial eta-squared ($\eta^2_p$) values \cite{cohen1973eta} for the factors were $0.462$ for \textit{Participant}, $0.241$ for \textit{Position}, and $0.165$ for \textit{Participant}$\times$\textit{Position}.
This can be interpreted as follows: (1) individual-to-individual differences are very large (0.462), and (2) there is a large effect of sensor position (0.241). Also, (3) a large interaction effect is present (0.165), 
suggesting that people differ in their optimal sensor positions.

To estimate the advantage of having a personalized optimal sensor position relative to the global optimum (center), we calculated the performance difference between the personal best and the 50\% condition. The averaged TP for the best positions was 6.034 bits/s, and the averaged TP at the 50\% position was 5.798 bits/s. This is roughly a 4.1\% throughput advantage (0.236 bits/s). 

\begin{figure}[t!]
	\centering
	\includegraphics[width=0.9\columnwidth]{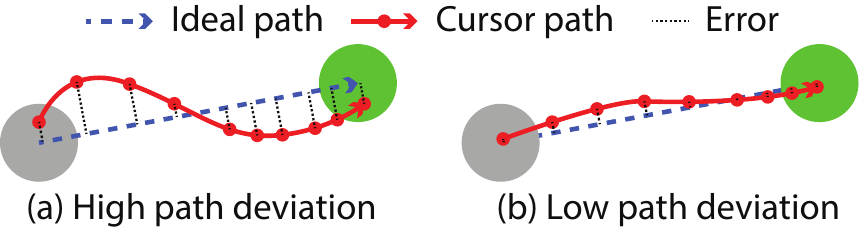}
	\caption{Path deviation, measured with mean absolute error ($MAE$) between cursor path and the ideal path, is used to measure variability in repeated acts of pointing. The ideal path is marked as a straight line connecting the origin and the target.}~\label{fig:path_dev} 
	\vspace{-2em}
\end{figure}

\textbf{Path deviation:}
To measure path deviation, we calculated mean absolute error (MAE). The error was defined as a distance between the committed cursor path and the ideal path (Figure \ref{fig:path_dev}). MAE for a path was calculated as ${\frac{1}{N}\sum_{i=1}^{N}|\textit{Error}_i|}$ where $N$ is all the points in the cursor trajectory and $\textit{Error}_i$ is the error at point $i$. A high path deviation implies a user was struggling to direct the cursor toward the target because the cursor movement diverged from the user's intentions.

\begin{figure}[t!]
	\centering
	\includegraphics[width=0.95\columnwidth]{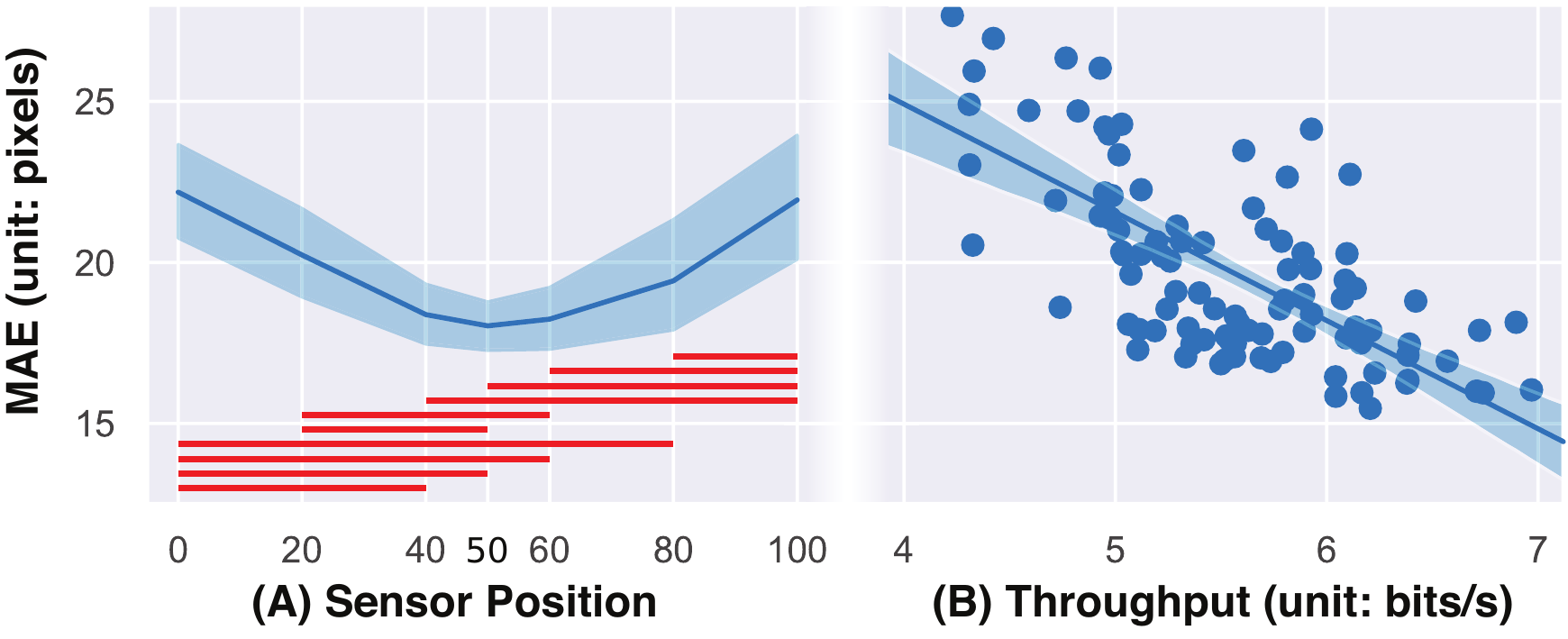}
	\caption{Sensor position affects path deviation in pointing: (A) MAE plotted for various sensor positions, the horizontal bars indicate a significant difference between positions; (B) linear regression showing a significant (\textit{p}<.001) negative association (\textit{r}=-.69) between TP and MAE. }~\label{fig:TP_vs_MAE}  
	\vspace{-1em}
\end{figure}

Figure \ref{fig:TP_vs_MAE}A shows the MAE values ranged 15.5--31.1 pixels (Mean 19.8, $SD$ 3.1) for different sensor positions.
 RM-ANOVA revealed a significant effect of sensor position on MAE ($F_{(3.13, 40.6)} = 19.02 $, $ p <.001 $, $ \eta^2_p = .594 $ with Greenhouse-Geisser correction). 
 The distribution of MAE had a quadratic form ($p<.001$).
 Bonferroni \emph{post-hoc} analyses showed higher values for the extremal positions (0\% and 100\%) in general; and 50\% and 60\% formed the best homogeneous subset.
 Compare to the 50\% position (MAE=18.0), 100\% position was worse by 19.9\% (MAE=21.6), and 0\% was worse by 22.5\% (MAE=22.1).
 This result is highly consistent with those of the TP analysis. 
 
\begin{figure}[tb!]
	\centering
	\includegraphics[width=0.85\columnwidth]{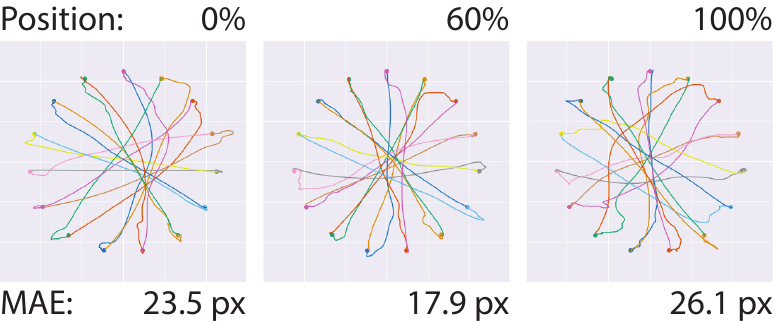}
	\caption{An example of high and low path deviations exhibited by P14 in three sessions ($D$=900, $W$=20, the optimal (60\%) and the extremal sensor positions). At the optimal position, the paths tend to be more straight and less deviates toward the target.}~\label{fig:path_dev_ex}
	\vspace{-1.5em}
\end{figure}

 In addition, we performed a linear regression analysis between TP and MAE, which revealed a strong negative correlation (see Figure \ref{fig:TP_vs_MAE}B). 
 This suggests that the performance drop may be attributable to greater path deviation. To conclude, when the sensor is placed in a suboptimal position, cursor trajectories vary more and are less under the user's control (see Figure \ref{fig:path_dev_ex}).
 
 As an alternative metric for measuring path deviation, we also calculated Root-Mean-Square Error (RMSE). The RMSE values ranged 18.7--40.3 pixels (Mean 24.0, $SD$ 4.0).
 RM-ANOVA revealed a significant effect of sensor position on RMSE ($F_{(2.9, 37.5)} = 16.96 $, $ p <.001 $, $ \eta^2_p = .566 $ with Greenhouse-Geisser correction). 
 The distribution of RMSE had a quadratic form ($p<.001$), centered at the 50\% position.
 Bonferroni \emph{post-hoc} analyses showed higher errors in the extremal positions (0\% and 100\%) than all others; however, no significant difference was found for the 20\%--80\% positions. Linear regression analysis exhibited a strong negative correlation ($r=-.70$) between TP and RMSE. The same conclusion is drawn as in the MAE analysis.
 
\subsection{Discussion}
In all the TP, MAE, and RMSE measurements, the data followed a quadratic distribution centered at the 50\% position. This suggests an advantage for the center position within our study setting.
The results obtained fly in the face of the strongly held belief in ``a dramatic advantage of the front sensor'' \cite{moggridge2007designing}, but aligned with the ``center of gravity'' hypothesis rather more. 
The sensor in the foremost position was almost as bad as the sensor in the rearmost location.

%% file: sec_calibration.tex
\begin{figure}[b!]
	\centering
	\includegraphics[width=0.99\columnwidth]{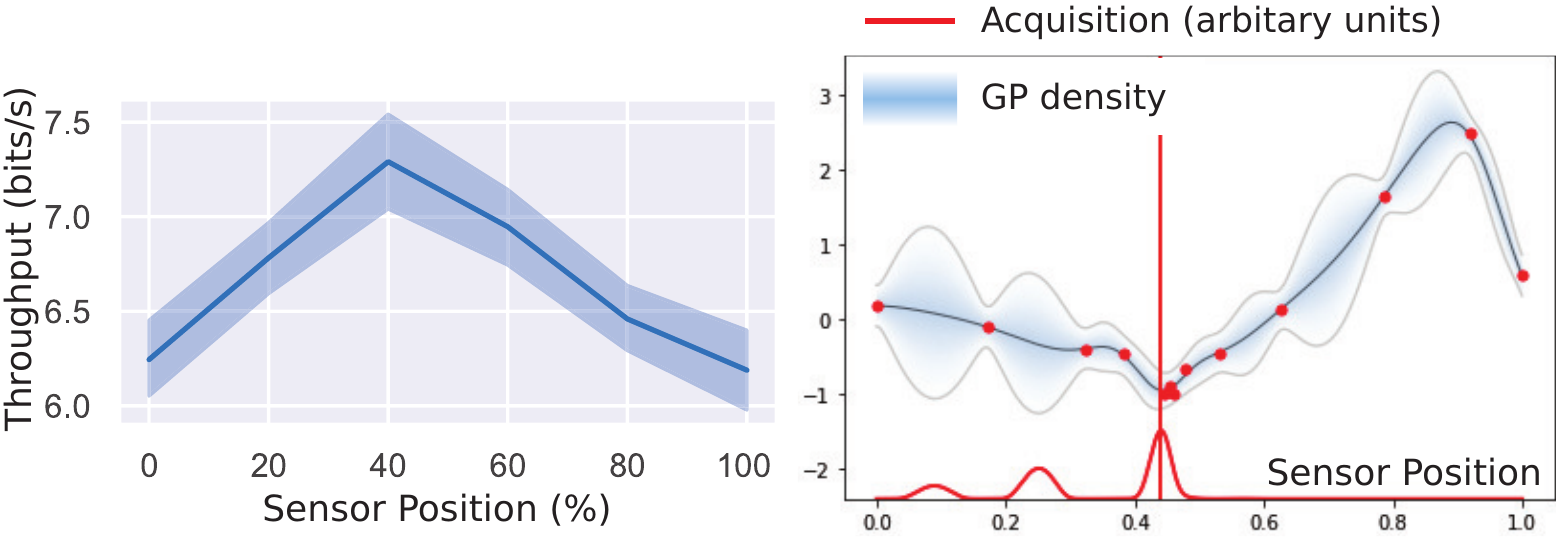}
	\caption{
		Sensor position can be individualized using a calibration procedure or in-application optimization.
		This figure demonstrates single-subject data on both. (Left) Calibration results in the multi-directional tapping test.
		(Right) Bayesian human-in-the-loop optimization using AimBooster. Both results converged to $p=40\%$ as the optimum.}~\label{fig:optimizer} 
\end{figure}

The virtual-sensor mouse permits software-side setting of the sensor position.
This can be exploited for improving the position to suit the individual or task.
We present two procedures and the related software\footnote{\url{https://userinterfaces.aalto.fi/mouse_sensor_position/}}, which take slightly different approaches. A single-subject demonstration of the two procedures is provided in Figure \ref{fig:optimizer}.

\textbf{The calibration task:} 
The virtual-sensor mouse can be calibrated for an individual in a one-shot manner employing the pointing task.
The entire calibration session takes about one hour, 
to guarantee enough repetitions for observing subtler differences between conditions.
We release software for carrying out the calibration with the virtual-sensor mouse.  
The calibration procedure follows the experimental method of the pointing study and is organized thus:
The software selects a sensor position randomly from the set ${20, 40, ..., 80\%}$,
then asks the user to perform the multi-directional tapping task for four minutes.
Extreme positions (0\% and 100\%) are removed since,
according to our study,
they are unlikely to offer the best settings.
Rest is enforced before moving on to the next position.
Confidence intervals are computed \emph{post hoc} for the throughput values, and path deviations obtained, like that reported in the experiment.
To choose the optimal sensor position for the user, we look at the highest-achieving conditions, and we choose the median $p$ if there is more than one within the confidence bounds.  
The procedure can be repeated for increased confidence in the optimal position, or test for the other sensor resolution.

\textbf{In-task optimization:} 
In addition to the controlled calibration procedure,
the virtual-sensor device can be exploited for human-in-the-loop optimization.
The position of the sensor can be optimized in any application wherein 
(1) a user aims toward a stationary target and selects by a click, (2) it requires repetitive target selections, and (3) clicks and cursor movement, including timestamps, can be logged. Most of the point-and-click games, for example, Whac-A-Mole, AimBooster, etc., are subject to be optimized.


An objective function can be formulated on the basis of the close correlation between throughput and path deviation, MAE or RMSE. We defined and used path deviation rate $PDR=\frac{\textrm{MAE}}{|Path|}$ or $\frac{\textrm{RMSE}}{|Path|}$, where $|Path|$ is the length of the ideal path. With this, we could relax the condition of prior knowledge about the target in throughput measurement. Divide by the path length will further compensate for the effect of varying movement amplitude. The ideal path is assumed to be a straight line between the start point (the beginning of the largest amplitude sub-movement \cite{lee2020autogain}) and the clicked point.
%

To facilitate deployment, we offer a Bayesian optimizer script in Python for sensor-position optimization \cite{shahriari2015taking}. It uses a sparse Gaussian Process (GP) as the proxy model to avoid choking on the high volume of data \cite{gpyopt2016}.
The proxy model maps sensor position $p$ to PDR, which it tries to minimize.
We recommend using expected improvement (EI) as the acquisition function;
which exploits our finding that most users' personal optima are somewhere near the center,
and that the form of the \emph{p}-to-deviation relationship is quadratic.  
The optimizer changes $p$ of the virtual mouse, and PDR is measured during a task. It takes about one minute for one sample to update the GP model. Hence, it should be called not mid-game but, for example, during breaks.

%% file: sec_discussion.tex
With this paper, 
we have presented two variable-sensor-position mouse constructions and shown that an accurate virtual mouse model can be achieved via a fusion of two sensors.
We described a user study in which we systematically manipulated the position of the sensor,
ascertaining that position does have a reliable effect on pointing performance.
Further, we showed that individuals differ in their optimal sensor position and, hence, calibrating the sensor yields additional benefits.

While we did not study the causal mechanisms underlying this effect, we hypothesize that they are attributable to musculoskeletal differences.
We believe that even small differences in rotation angles, posture, and limb lengths would carry over to this effect.
Alternatively, it may be due to the cognitive mechanism by which people abstract their hand (and a device) into a point.
For example, 
the user's internal model may prefer the center of the mouse as its reference point when planning pointing movements. 
This mechanism may also explain the path deviation being minimal at the center sensor position during the pointing process.

To the best of our knowledge, sensor placement has never been examined systematically before. 
Our work opens up opportunities to understand differences among gamers, one aspect of which may be trivially attributable to the mouse designs chosen. 
While HCI research has focused on pointer-facilitation techniques for translational motion, biomechanical factors that affect rotational motion, such as coordinate disturbance and joint rotation, need more attention.
We foresee the data and hardware constructions that we release facilitating such efforts.

We can identify several opportunities to build on this work. Our study was limited to a pointing task, and it will be important to reproduce these effects in other tasks, such as tracing or drawing.
We believe that the effect will persist and may even be amplified in fast motions in which users rely more on ballistic motions.  
Also, our experimental task was relatively short. The effect should be replicated in a longitudinal study, with longer training times.
In addition, while our study considered only vertical changes in sensor position, 
the horizontal location too may affect the motion. 
This should be studied further.
We used the familiar Logitech design only, and it would be important to assess whether the position effect interacts with the shape and weight of the device or the CD gain functions.
We also expect the virtual sensor and the optimization techniques will be able to minimize the sense of incompatibility when using two or more mice by one person, for example, at work and at home. 

The experiment had certain limitations. 
Firstly, potential interaction between shape, sensor resolution, and sensor position was not examined. The preferred sensor position may change at lower or higher CPI values or with different forms, although we find this unlikely in light of the lack of an effect in the displacement study. 
Secondly, the participants may have had a bias in favor of the center position, produced by daily use.
To mitigate this, we provided a short but extensive (500+ clicks) training for every block before actual testing.
One evidence could be P13 who exhibited optimal position at 80\% (see on Figure~\ref{fig:throughput}). At least for P13, the training was enough to overcome the existing bias from the centered sensor of the everyday mouse.
However, the effectiveness of the training, in general, was not evaluated.
Repeating the study with differently biased people (e.g., everyday mice with front sensors) could be an immediate follow-up investigation.
%
Lastly, the calibration was demonstrated only with a single user. The result suggests preliminary evidence that the method will likely work. Large-scale deployment and examination will strengthen the validity of the calibration process.

\section{Conclusion}
Our work demonstrates that the choice of displacement sensor position affects pointing performance.
While the study result indicates that a central position is the best compromise,
performance can be further improved with a virtual-sensor device that allows personalizing the position.
Our data suggests that an additional improvement of 4\% is achievable with the personalized optimization.
This may seem small, but for top-competitors such as pro gamers, even a small improvement could contribute.
The virtual sensor is inexpensive to implement and would increase the cost of a high-end mouse only marginally.
We expect the findings to be of interest to HCI researchers, engineers, gamers, computer graphic designers, and anyone else engaged in performance-oriented mouse use.

\section{Acknowledgment and Open Science}
This work has been funded by the European Research Council (ERC) under the European Union's Horizon 2020 research and innovation programme (grant no. 637991), and Korea Creative Content Agency (award no. R2019020010). We thank Marko Repo for his great assistance in carrying out the experiments, and Francesco Verdoja and Jens Lundell for their help in the robotic arm experiment.
All data and code generated in the study are released at \url{https://userinterfaces.aalto.fi/mouse_sensor_position/}. The page reports 3D models of the mouse apparatus, circuit schematics, firmware, all raw and processed data reported here, as well as the scripts for data analysis and the calibration and optimization procedures.

%% file: main.bbl

\begin{thebibliography}{00}


\ifx \showCODEN    \undefined \def \showCODEN     #1{\unskip}     \fi
\ifx \showDOI      \undefined \def \showDOI       #1{{\tt DOI:}\penalty0{#1}\ }
  \fi
\ifx \showISBNx    \undefined \def \showISBNx     #1{\unskip}     \fi
\ifx \showISBNxiii \undefined \def \showISBNxiii  #1{\unskip}     \fi
\ifx \showISSN     \undefined \def \showISSN      #1{\unskip}     \fi
\ifx \showLCCN     \undefined \def \showLCCN      #1{\unskip}     \fi
\ifx \shownote     \undefined \def \shownote      #1{#1}          \fi
\ifx \showarticletitle \undefined \def \showarticletitle #1{#1}   \fi
\ifx \showURL      \undefined \def \showURL       #1{#1}          \fi

\bibitem{logitech2008}
 2008.
\newblock {Mice That Didn't Leave the Logitech Labs}.
\newblock
  \url{https://blog.logitech.com/2008/12/03/one-billion-logitech-mice/},
  \url{http://www.logitech.com/pub/onebillion/multimedia/mice_that_didnt_make_it.pdf}.
    (2008).
\newblock
\newblock
\shownote{(Accessed on Jan 03, 2019).}


\bibitem{myth_esr}
 2013.
\newblock ESR - New gaming mice (vps-)invention and more.
\newblock
  \url{http://www.esreality.com/post/2410040/new-gaming-mice-vps-invention-and-more/}.
    (2013).
\newblock
\newblock
\shownote{(Accessed on Jan 03, 2020).}


\bibitem{myth_oc1}
 2014.
\newblock The Importance of Sensor Positioning - Overclock.net - An
  Overclocking Community.
\newblock
  \url{https://www.overclock.net/forum/375-mice/1522415-importance-sensor-positioning.html}.
    (2014).
\newblock
\newblock
\shownote{(Accessed on Nov 19, 2019).}


\bibitem{myth_geekhack}
 2018.
\newblock Sensor location on mouse.
\newblock \url{https://geekhack.org/index.php?topic=96589.0}.   (2018).
\newblock
\newblock
\shownote{(Accessed on Jan 03, 2020).}


\bibitem{myth_reddit}
 2019.
\newblock Does nobody care about sensor placement? : MouseReview.
\newblock
  \url{https://www.reddit.com/r/MouseReview/comments/atyot7/does_nobody_care_about_sensor_placement/}.
    (2019).
\newblock
\newblock
\shownote{(Accessed on Jan 03, 2020).}


\bibitem{aceituno2013low}
{Jonathan Aceituno}, {G\'{e}ry Casiez}, {and} {Nicolas Roussel}. 2013.
\newblock \showarticletitle{How low can you go? human limits in small
  unidirectional mouse movements}. In {\em Proceedings of the SIGCHI Conference
  on Human Factors in Computing Systems} {\em (CHI '13)}. Association for
  Computing Machinery, New York, NY, USA, 1383--1386.
\newblock
\showISBNx{9781450318990}
\showDOI{%
\url{http://dx.doi.org/10.1145/2470654.2466182}}


\bibitem{almeida2006supporting}
{Rodrigo Almeida} {and} {Pierre Cubaud}. 2006.
\newblock \showarticletitle{Supporting 3D window manipulation with a yawing
  mouse}. In {\em Proceedings of the 4th Nordic Conference on Human-Computer
  Interaction: Changing Roles} {\em (NordiCHI '06)}. Association for Computing
  Machinery, New York, NY, USA, 477--480.
\newblock
\showISBNx{1595933255}
\showDOI{%
\url{http://dx.doi.org/10.1145/1182475.1182541}}


\bibitem{Apperley2013}
{Mark Apperley} {and} {Bill Rogers}. 2013.
\newblock \showarticletitle{{The orienting mouse: An input device with attitude
  (Wokring paper)}}.
\newblock  (Aug. 2013).
\newblock
\showURL{%
\url{https://hdl.handle.net/10289/8194}}


\bibitem{gpyopt2016}
{The~GPyOpt authors}. 2016.
\newblock GPyOpt: A Bayesian Optimization framework in Python.
\newblock \url{http://github.com/SheffieldML/GPyOpt}.   (2016).
\newblock


\bibitem{berard2011humanLimit}
{Fran\c{c}ois B{\'e}rard}, {Guangyu Wang}, {and} {Jeremy~R. Cooperstock}. 2011.
\newblock \showarticletitle{On the limits of the human motor control precision:
  the search for a device's human resolution}. In {\em Proceedings of the 13th
  IFIP TC 13 International Conference on Human-computer Interaction - Volume
  Part II} {\em (INTERACT'11)}. Springer-Verlag, Berlin, Heidelberg, 107--122.
\newblock
\showISBNx{978-3-642-23770-6}
\showURL{%
\url{http://dl.acm.org/citation.cfm?id=2042118.2042131}}


\bibitem{burgess1999wrist}
{R Burgess-Limerick}, {J Shemmell}, {R Scadden}, {and} {A Plooy}. 1999.
\newblock \showarticletitle{Wrist posture during computer pointing device use}.
\newblock {\em Clinical Biomechanics\/} {14}, 4 (1999), 280 -- 286.
\newblock
\showISSN{0268-0033}
\showDOI{%
\url{http://dx.doi.org/10.1016/S0268-0033(98)90093-6}}


\bibitem{cabeccas2010friction}
{Jos{\'e}~Miquel Cabe{\c{c}}as}. 2010.
\newblock \showarticletitle{The friction force mouse-pad and the forearm
  muscles efforts}.
\newblock {\em The Ergonomics Open Journal\/}  {3} (2010), 1--13.
\newblock


\bibitem{casiez2011no}
{G\'{e}ry Casiez} {and} {Nicolas Roussel}. 2011.
\newblock \showarticletitle{No more bricolage! Methods and tools to
  characterize, replicate and compare pointing transfer functions}. In {\em
  Proceedings of the 24th Annual ACM Symposium on User Interface Software and
  Technology} {\em (UIST '11)}. Association for Computing Machinery, New York,
  NY, USA, 603--614.
\newblock
\showISBNx{9781450307161}
\showDOI{%
\url{http://dx.doi.org/10.1145/2047196.2047276}}


\bibitem{casiez08}
{G\'{e}ry Casiez}, {Daniel Vogel}, {Ravin Balakrishnan}, {and} {Andy Cockburn}.
  2008.
\newblock \showarticletitle{The impact of control-display gain on user
  performance in pointing tasks}.
\newblock {\em Human-Computer Interaction\/} {23}, 3 (2008), 215--250.
\newblock
\showDOI{%
\url{http://dx.doi.org/10.1080/07370020802278163}}


\bibitem{chaparro2000range}
{Alex Chaparro}, {Michael Rogers}, {Jeffrey Fernandez}, {Mike Bohan},
  {Choi~Sang Dae}, {and} {Laszlo Stumpfhauser}. 2000.
\newblock \showarticletitle{Range of motion of the wrist: implications for
  designing computer input devices for the elderly}.
\newblock {\em Disability and Rehabilitation\/} {22}, 13-14 (2000), 633--637.
\newblock
\showDOI{%
\url{http://dx.doi.org/10.1080/09638280050138313}}
\newblock
\shownote{PMID: 11052213.}


\bibitem{chen2012weight}
{Han-Ming Chen}, {Chang-Sian Lee}, {and} {Chih-Hsiu Cheng}. 2012.
\newblock \showarticletitle{The weight of computer mouse affects the wrist
  motion and forearm muscle activity during fast operation speed task}.
\newblock {\em European Journal of Applied Physiology\/} {112}, 6 (01 Jun
  2012), 2205--2212.
\newblock
\showISSN{1439-6327}
\showDOI{%
\url{http://dx.doi.org/10.1007/s00421-011-2198-3}}


\bibitem{Chen2007ey}
{Han-Ming Chen} {and} {Chun-Tong Leung}. 2007.
\newblock \showarticletitle{The effect on forearm and shoulder muscle activity
  in using different slanted computer mice}.
\newblock {\em Clinical Biomechanics\/} {22}, 5 (2007), 518 -- 523.
\newblock
\showISSN{0268-0033}
\showDOI{%
\url{http://dx.doi.org/10.1016/j.clinbiomech.2007.01.006}}


\bibitem{cohen1973eta}
{Jacob Cohen}. 1973.
\newblock \showarticletitle{Eta-squared and partial eta-squared in fixed factor
  anova designs}.
\newblock {\em Educational and Psychological Measurement\/} {33}, 1 (1973),
  107--112.
\newblock
\showDOI{%
\url{http://dx.doi.org/10.1177/001316447303300111}}


\bibitem{dehghan2013designing}
{Naser Dehghan}, {Alireza Choobineh}, {Mohsen Razeghi}, {Jafar Hasanzadeh},
  {and} {Moslem Irandoost}. 2013.
\newblock \showarticletitle{Designing a new computer mouse and evaluating some
  of its functional parameters}.
\newblock {\em Journal of research in health sciences\/} {14}, 2 (2013),
  132--135.
\newblock


\bibitem{engelbart1970xy}
{Douglas~C Engelbart}. 1970.
\newblock Xy position indicator for a display system.
\newblock   (Nov.~17 1970).
\newblock
\newblock
\shownote{US Patent 3,541,541.}


\bibitem{fallman20073dof}
{Daniel Fallman}, {Anneli Mikaelsson}, {and} {Bj{\"o}rn Yttergren}. 2007.
\newblock \showarticletitle{The design of a computer mouse providing three
  degrees of freedom}. In {\em Human-Computer Interaction. Interaction
  Platforms and Techniques}, {Julie~A. Jacko} (Ed.). Springer Berlin
  Heidelberg, Berlin, Heidelberg, 53--62.
\newblock
\showISBNx{978-3-540-73107-8}
\showDOI{%
\url{http://dx.doi.org/10.1007/978-3-540-73107-8_7}}


\bibitem{fitts1954information}
{Paul~M Fitts}. 1954.
\newblock \showarticletitle{The information capacity of the human motor system
  in controlling the amplitude of movement.}
\newblock {\em Journal of experimental psychology\/} {47}, 6 (1954), 381.
\newblock
\showDOI{%
\url{http://dx.doi.org/10.1037/h0055392}}


\bibitem{gustafsson2003computer}
{Ewa Gustafsson} {and} {Mats Hagberg}. 2003.
\newblock \showarticletitle{Computer mouse use in two different hand positions:
  exposure, comfort, exertion and productivity}.
\newblock {\em Applied ergonomics\/} {34}, 2 (2003), 107--113.
\newblock
\showDOI{%
\url{http://dx.doi.org/10.1016/S0003-6870(03)00005-X}}


\bibitem{hannagan2007twistmouse}
{Jacqui Hannagan}. 2007.
\newblock {\em TwistMouse for simultaneous translation and rotation}.
\newblock Ph.D. Dissertation.
\newblock
\showURL{%
\url{http://hdl.handle.net/10523/1176}}


\bibitem{Hedge2010ep}
{Alan Hedge}, {David Feathers}, {and} {Kimberly Rollings}. 2010.
\newblock \showarticletitle{{Ergonomic comparison of slanted and vertical
  computer mouse designs}}. In {\em Proceedings of the Human Factors and
  Ergonomics Society}. Cornell University, Ithaca, United States, 561--565.
\newblock
\showDOI{%
\url{http://dx.doi.org/10.1177/154193121005400604}}


\bibitem{houwink2009providing}
{Annemieke Houwink}, {Karen~M Oude~Hengel}, {Dan Odell}, {and} {Jack~T
  Dennerlein}. 2009.
\newblock \showarticletitle{Providing training enhances the biomechanical
  improvements of an alternative computer mouse design}.
\newblock {\em Human Factors\/} {51}, 1 (2009), 46--55.
\newblock
\showDOI{%
\url{http://dx.doi.org/10.1177/0018720808329843}}


\bibitem{Hughes2012}
{Erin~E Hughes} {and} {Peter~W Johnson}. 2012.
\newblock \showarticletitle{{Children computer mouse use and anthropometry}}.
  In {\em Work}. University of Washington, Seattle, Seattle, United States, IOS
  Press, 846--850.
\newblock
\showDOI{%
\url{http://dx.doi.org/10.3233/WOR-2012-0252-846}}


\bibitem{isokoski2002speed}
{Poika Isokoski} {and} {Roope Raisamo}. 2002.
\newblock \showarticletitle{Speed-accuracy measures in a population of six
  mice}. In {\em Proc. APCHI2002: 5th Asia Pacific Conference on Computer Human
  Interaction}. Science Press, 765--777.
\newblock


\bibitem{jensen1998job}
{Chris Jensen}, {Vilhelm Borg}, {Lotte Finsen}, {Klaus Hansen}, {Birgit
  Juul-Kristensen}, {and} {Hanne Christensen}. 1998.
\newblock \showarticletitle{Job demands, muscle activity and musculoskeletal
  symptoms in relation to work with the computer mouse}.
\newblock {\em Scandinavian Journal of Work, Environment \& Health\/} {24}, 5
  (1998), 418--424.
\newblock
\showISSN{03553140, 1795990X}
\showURL{%
\url{http://www.jstor.org/stable/40966801}}


\bibitem{keir1999effects}
{Peter~J Keir}, {Joel~M Bach}, {and} {David Rempel}. 1999.
\newblock \showarticletitle{Effects of computer mouse design and task on carpal
  tunnel pressure}.
\newblock {\em Ergonomics\/} {42}, 10 (1999), 1350--1360.
\newblock
\showDOI{%
\url{http://dx.doi.org/10.1080/001401399184992}}


\bibitem{kim2008inflatable}
{Seoktae Kim}, {Hyunjung Kim}, {Boram Lee}, {Tek-Jin Nam}, {and} {Woohun Lee}.
  2008.
\newblock \showarticletitle{Inflatable mouse: volume-adjustable mouse with
  air-pressure-sensitive input and haptic feedback}. In {\em Proceedings of the
  SIGCHI Conference on Human Factors in Computing Systems} {\em (CHI '08)}.
  Association for Computing Machinery, New York, NY, USA, 211--224.
\newblock
\showISBNx{9781605580111}
\showDOI{%
\url{http://dx.doi.org/10.1145/1357054.1357090}}


\bibitem{Lee2015}
{Byungjoo Lee} {and} {Hyunwoo Bang}. 2015.
\newblock \showarticletitle{A mouse with two optical sensors that eliminates
  coordinate disturbance during skilled strokes}.
\newblock {\em Human--Computer Interaction\/} {30}, 2 (2015), 122--155.
\newblock
\showDOI{%
\url{http://dx.doi.org/10.1080/07370024.2014.894888}}


\bibitem{lee2020autogain}
{Byungjoo Lee}, {Mathieu Nancel}, {Sunjun Kim}, {and} {Antti Oulasvirta}. 2020.
\newblock \showarticletitle{AutoGain: Gain Function Adaptation with Submovement
  Efficiency Optimization}. In {\em Proceedings of the SIGCHI Conference on
  Human Factors in Computing Systems} {\em (CHI '20)}. ACM, New York, NY, USA.
\newblock
\showDOI{%
\url{http://dx.doi.org/10.1145/3313831.3376244}}


\bibitem{lee2007alternative}
{David~L Lee}, {Jacob Fleisher}, {Hugh~E McLoone}, {Kentaro Kotani}, {and}
  {Jack~T Dennerlein}. 2007.
\newblock \showarticletitle{Alternative computer mouse design and testing to
  reduce finger extensor muscle activity during mouse use}.
\newblock {\em Human Factors\/} {49}, 4 (2007), 573--584.
\newblock
\showDOI{%
\url{http://dx.doi.org/10.1518/001872007X215665}}


\bibitem{mackenzie1992fitts}
{I~Scott MacKenzie}. 1992.
\newblock \showarticletitle{Fitts' law as a research and design tool in
  human-computer interaction}.
\newblock {\em Human-computer interaction\/} {7}, 1 (1992), 91--139.
\newblock
\showDOI{%
\url{http://dx.doi.org/10.1207/s15327051hci0701_3}}


\bibitem{mackenzie2015user}
{I.~Scott MacKenzie}. 2015.
\newblock \showarticletitle{User studies and usability evaluations: from
  research to products}. In {\em Proceedings of the 41st Graphics Interface
  Conference} {\em (GI '15)}. Canadian Information Processing Society, CAN,
  1--8.
\newblock
\showISBNx{9780994786807}


\bibitem{MacKenzie1997}
{I.~Scott MacKenzie}, {R.~William Soukoreff}, {and} {Chris Pal}. 1997.
\newblock \showarticletitle{A two-ball mouse affords three degrees of freedom}.
  In {\em CHI '97 Extended Abstracts on Human Factors in Computing Systems}
  {\em (CHI EA '97)}. Association for Computing Machinery, New York, NY, USA,
  303--304.
\newblock
\showISBNx{0897919262}
\showDOI{%
\url{http://dx.doi.org/10.1145/1120212.1120405}}


\bibitem{moggridge2007designing}
{Bill Moggridge} {and} {Bill Atkinson}. 2007.
\newblock {\em Designing interactions}. Vol.~17.
\newblock MIT press Cambridge, MA.
\newblock


\bibitem{odell2007evaluation}
{Daniel~L Odell} {and} {Peter~W Johnson}. 2007.
\newblock \showarticletitle{Evaluation of a mouse designed to improve posture
  and comfort}. In {\em Proceedings of the 2007 Work with Computing Systems
  Conference-International Ergonomics Association}.
\newblock


\bibitem{poston2007computer}
{Timothy Poston} {and} {Manohar Srikanth}. 2007.
\newblock Computer input device enabling three degrees of freedom and related
  input and feedback methods.
\newblock   (June~28 2007).
\newblock
\newblock
\shownote{US Patent App. 11/616,653.}


\bibitem{quemelo2013biomechanics}
{Paulo~RV Quemelo} {and} {Edgar~Ramos Vieira}. 2013.
\newblock \showarticletitle{Biomechanics and performance when using a standard
  and a vertical computer mouse}.
\newblock {\em Ergonomics\/} {56}, 8 (2013), 1336--1344.
\newblock
\showDOI{%
\url{http://dx.doi.org/10.1080/00140139.2013.805251}}


\bibitem{roussel2012subpixel}
{Nicolas Roussel}, {G{\'e}ry Casiez}, {Jonathan Aceituno}, {and} {Daniel
  Vogel}. 2012.
\newblock \showarticletitle{Giving a hand to the eyes: leveraging input
  accuracy for subpixel interaction}. In {\em Proceedings of the 25th Annual
  ACM Symposium on User Interface Software and Technology} {\em (UIST '12)}.
  ACM, New York, NY, USA, 351--358.
\newblock
\showISBNx{978-1-4503-1580-7}
\showDOI{%
\url{http://dx.doi.org/10.1145/2380116.2380162}}


\bibitem{shahriari2015taking}
{Bobak Shahriari}, {Kevin Swersky}, {Ziyu Wang}, {Ryan~P Adams}, {and} {Nando
  De~Freitas}. 2015.
\newblock \showarticletitle{Taking the human out of the loop: A review of
  Bayesian optimization}.
\newblock {\it Proc. IEEE} {104}, 1 (2015), 148--175.
\newblock
\showDOI{%
\url{http://dx.doi.org/10.1109/JPROC.2015.2494218}}


\bibitem{shoemaker2012two}
{Garth Shoemaker}, {Takayuki Tsukitani}, {Yoshifumi Kitamura}, {and}
  {Kellogg~S. Booth}. 2012.
\newblock \showarticletitle{Two-part models capture the impact of gain on
  pointing performance}.
\newblock {\em ACM Transactions on Computer-Human Interaction (TOCHI)\/} {19},
  4, Article 28 (Dec. 2012), 34 pages.
\newblock
\showISSN{1073-0516}
\showDOI{%
\url{http://dx.doi.org/10.1145/2395131.2395135}}


\bibitem{soukoreff2004towards}
{R~William Soukoreff} {and} {I~Scott MacKenzie}. 2004.
\newblock \showarticletitle{Towards a standard for pointing device evaluation,
  perspectives on 27 years of Fitts' law research in HCI}.
\newblock {\em International journal of human-computer studies\/} {61}, 6
  (2004), 751--789.
\newblock
\showDOI{%
\url{http://dx.doi.org/10.1016/j.ijhcs.2004.09.001}}


\bibitem{tang2010adaptive}
{Sheng~Kai Tang} {and} {Wen~Yen Tang}. 2010.
\newblock \showarticletitle{Adaptive mouse: a deformable computer mouse
  achieving form-function synchronization}. In {\em CHI '10 Extended Abstracts
  on Human Factors in Computing Systems} {\em (CHI EA '10)}. Association for
  Computing Machinery, New York, NY, USA, 2785--2792.
\newblock
\showISBNx{9781605589305}
\showDOI{%
\url{http://dx.doi.org/10.1145/1753846.1753864}}


\bibitem{verplank1989microsoft}
{Bill Verplank} {and} {Kate Oliver}. 1989.
\newblock \showarticletitle{Microsoft mouse: testing for redesign}. In {\em
  Proceedings of INTERFACE'89, The Sixth Symposium on Human Factors and
  Industrial Design in Consumer Products}. 257--261.
\newblock
\showURL{%
\url{http://www.billverplank.com/MouseTests/}}
\newblock
\shownote{(Accessed on Jan 13, 2020).}


\bibitem{wahlstrom2000differences}
{Jens Wahlstrom}, {Joakim Svensson}, {Mats Hagberg}, {and} {Peter~W Johnson}.
  2000.
\newblock \showarticletitle{Differences between work methods and gender in
  computer mouse use}.
\newblock {\em Scandinavian journal of work, environment \& health\/} {26}, 5
  (2000), 390--397.
\newblock


\bibitem{williams1949balanced}
{EJ Williams}. 1949.
\newblock \showarticletitle{Experimental designs balanced for the estimation of
  residual effects of treatments}.
\newblock {\em Australian Journal of Chemistry\/} {2}, 2 (1949), 149--168.
\newblock
\showDOI{%
\url{http://dx.doi.org/10.1071/CH9490149}}


\bibitem{anglemouse}
{Jacob~O. Wobbrock}, {James Fogarty}, {Shih-Yen~(Sean) Liu}, {Shunichi Kimuro},
  {and} {Susumu Harada}. 2009.
\newblock \showarticletitle{The angle mouse: target-agnostic dynamic gain
  adjustment based on angular deviation}. In {\em Proceedings of the SIGCHI
  Conference on Human Factors in Computing Systems} {\em (CHI '09)}. ACM, New
  York, NY, USA, 1401--1410.
\newblock
\showISBNx{978-1-60558-246-7}
\showDOI{%
\url{http://dx.doi.org/10.1145/1518701.1518912}}


\bibitem{wobbrock2011effects}
{Jacob~O. Wobbrock}, {Kristen Shinohara}, {and} {Alex Jansen}. 2011.
\newblock \showarticletitle{The effects of task dimensionality, endpoint
  deviation, throughput calculation, and experiment design on pointing measures
  and models}. In {\em Proceedings of the SIGCHI Conference on Human Factors in
  Computing Systems} {\em (CHI '11)}. Association for Computing Machinery, New
  York, NY, USA, 1639--1648.
\newblock
\showISBNx{9781450302289}
\showDOI{%
\url{http://dx.doi.org/10.1145/1978942.1979181}}


\end{thebibliography}
